\documentclass[11pt]{article}

\usepackage{hyperref}
\hypersetup{colorlinks=false}

\usepackage{color}
\usepackage{latexsym}
\usepackage{amssymb}
\usepackage{amsmath}
\usepackage{graphicx}
\usepackage{subfig}
\usepackage{hhline}

\newtheorem{Theorem}{Theorem}[part]

\newtheorem{Proposition}{Proposition}[part]

\newtheorem{Remark}{Remark}[part]

\def \Frac{\displaystyle\frac}

\def \N{\mathbb{N}}
\def \R{\mathbb{R}}

\def \Z{\mathbb{Z}}
\def \E{\mathbb{E}}
\def \F{\mathbb{F}}

\def\Y{\mathbb{Y}}
\def \P{\mathbb{P}}

\def \T{\mathbb{T}}

\def\Pb{{\bf  P}}
\def\Eb{{\bf  E}}

\def \Ac{{\cal A}}

\def \Fc{{\cal F}}

\def \Ic{{\cal I}}

\def \Lc{{\cal L}}
\def \Pc{{\cal P}}

\def \Mc{{\cal M}}

\def \Sc{{\cal S}}
\def \Tc{{\cal T}}

\def \eps{\varepsilon}

\def \ep{\hbox{ }\hfill$\Box$}

\def\Dt#1{\Frac{\partial #1}{\partial t}}

\def\reff#1{{\rm(\ref{#1})}}

\def\beqs{\begin{eqnarray*}}
\def\enqs{\end{eqnarray*}}
\def\beq{\begin{eqnarray}}
\def\enq{\end{eqnarray}}

\begin{document}

\title{Optimal high-frequency trading in a pro-rata microstructure with predictive information}

\author{Fabien GUILBAUD
             \\\small Laboratoire de Probabilit\'es et
             \\\small  Mod\`eles Al\'eatoires
             \\\small  CNRS, UMR 7599
             \\\small  Universit\'e Paris 7 Diderot
             \\\small  fabien.guilbaud@gmail.com
             \and
              Huy\^en PHAM
             \\\small  Laboratoire de Probabilit\'es et
             \\\small  Mod\`eles Al\'eatoires
             \\\small  CNRS, UMR 7599
             \\\small  Universit\'e Paris 7 Diderot
             \\\small  pham@math.univ-paris-diderot.fr
             \\\small   and CREST-ENSAE 
             }


\maketitle

\begin{abstract}
We propose a framework to study optimal trading policies in a one-tick pro-rata limit order book, as typically arises in short-term interest rate futures contracts. The high-frequency trader has the choice to trade via market orders or limit orders, which are represented respectively by impulse controls and regular controls. We model and discuss the consequences of the two main features of this particular microstructure: first, the limit orders sent by the high frequency trader are only partially executed, and therefore she has no control on the executed quantity. 
For this purpose, cumulative executed volumes are modelled  by compound Poisson  processes. 
Second, the high frequency trader faces the overtrading risk, which is the risk of brutal variations in her inventory. 
The consequences of this risk are investigated in the context of optimal liquidation.

The optimal trading problem is studied by stochastic control and dynamic progra\-mming methods, which lead to a characterization of the value function in terms of an integro quasi-variational inequality. We then provide the associated numerical resolution procedure, and convergence of this computational scheme is proved. Next, we examine several situations where we can on one hand simplify the numerical procedure by reducing the number of state variables, and on the other hand focus on specific cases of practical interest. We examine both a market making problem and a best execution problem in the case where the mid-price process is a martingale. We also detail a high frequency trading strategy in the case where a (predictive) directional information on the mid-price is available. Each of the resulting strategies are illustrated by numerical tests.
\end{abstract}

\vspace{3mm}


\newpage

\section{Introduction}


In most of modern public security markets, the price formation process, or price discovery, results from  competition between several market agents that take part in a public auction. In particular, day trading sessions, which are also called continuous trading phases, consist of continuous double auctions. In these situations, liquidity providers\footnote{In this paper, we call liquidity provider any investor that currently trades with limit orders} continuously set bid and ask prices for the considered security, and the marketplace publicly displays a (possibly partial) information about these bid and ask prices, along with transactions prices. The action of continuously providing bid and ask quotes during day trading sessions is called \textit{market making}, and this role was tradionnally performed by specialist firms. However, due to the recent increased availability of electronic trading technologies, as well as regulatory changes, a large range of investors are now able to implement such market making strategies. These strategies are part of the broader category of high frequency trading (HFT) strategies, which are characterized by the fact that they facilitate a larger number of orders being sent to the market per unit of time. HFT takes place in the continuous trading phase, and therefore in the double continuous auction context, but actual mechanisms that implement this general continuous double auction set-up directly influence the price formation process and, as a consequence, HFT strategies.


In this work, we  shall focus on the case where the continuous double auction is implemented by a limit order book (LOB), operated under the pro-rata microstructure\footnote{This microstructure differs from the price/time microstructure that can be encountered on most cash equity markets, see \cite{guipha11}. }, see \cite{jankab07} and \cite{aik06}. This microstructure can be encountered on some derivatives markets, and especially in short-term interest rate (STIR) futures markets, also known as financial futures, traded e.g. on LIFFE (London International Financial Futures and options Exchange) or on CME (Chicago Mercantile Exchange). We will describe this microstructure in depth in Section 2, but the general mechanism of this microstructure is as follows: an incoming market order is  dispatched on all active limit orders at the best price, with each limit order contributing to execution in proportion to its volume. In particular, we will discuss the two main consequences of this microstructure on HFT strategies which are the oversizing of the best priced slices of the LOB and the overtrading risk.

Our main goal is to construct an HFT strategy, by means of optimal stochastic control, that targets the pro-rata microstructure. We allow both limit orders and market orders in this HFT strategy, modelled respectively as continuous and impulse controls, due to considerations about direct trading costs. From a modelling point of view, the key novelty is that we take into account partial execution for limit orders, which is crucial in the pro-rata case. For this purpose we introduce a Poissonian model for trades processes, that can be fitted to a large class of real-world execution processes, since we make few assumptions about the distributions of execution volumes. From a practical trading point of view, we allow the HFT to input predictive information about price evolution into the strategy, so that our algorithm can be seen as an information-driven HFT strategy (this situation is sometimes called HFT with superior information, see \cite{carjairic11}). We derive the dynamic programming equation corresponding to this mixed impulse/regular control problem. Moreover, we are able to reduce the number of relevant state variables to one in two situations of practical interest: first, in the simple case where the mid-price is a martingale, and second, in the case where the mid-price is a L\'evy process, in particular when the HFT has predictive information on price trend, in line with recent studies \cite{conlar10}. 
We provide a computational algorithm for the resolution of the dynamic programming equation, and prove the convergence of this scheme. We   illustrate numerically the behavior of the strategy and perform a simulated data benchmarked backtest.

High-frequency trading has recently received sustained academic interest. The reference work for inventory-based high frequency trading is Avellaneda and Stoikov (2008) \cite{avesto08}. The authors present the HFT problem as an inventory management problem and define inventory risk as the risk of holding a non-zero position in a risky asset. They also provide a closed-form approximate solution in a stylized market model where the controls are continuous. Several works are available that describe optimal strategies for HFT on cash equities or foreign exchange, e.g. \cite{kuhstr10},  
\cite{gueferleh11}, \cite{guipha11} or \cite{ver11}. Gu\'eant, Tapia and Lehalle (\cite{gueferleh11}) provide extensive analytical treatment of the Avellaneda and Stoikov model. Veraart (\cite{ver11})  includes market orders (that are modelled as impulse controls) as well as limit order in the context of FX trading. Guilbaud and Pham (\cite{guipha11}) study  market/limit orders HFT strategies on stocks with a focus on  the price/time microstructure and the bid/ask spread modelling. More recently, Cartea, Jaimungal and Ricci (\cite{carjairic11}) consider a HFT strategy that takes into account influence of trades on the LOB, and give the HFT superior information about the security price evolution. A growing literature is dedicated to modelling the dynamics of the limit order book itself, and its consequences for the price formation process. A popular approach is the Poisson Limit Order Book model as in  Cont and de Larrard (\cite{conlar10}). These authors  are able to retrieve a predictive information on price behavior (together with other LOB features) based on the current state of the order book. Finally, in empirical literature, much work is available for cash equities e.g. \cite{LOBsurvey}, but very few is dedicated to markets operating under pro-rata microstructure. We would like to mention the  work by Field and Large (\cite{fielar08}), which provides a detailed empirical description of pro-rata microstructure.

This paper is organized as follows: in Section 2, we detail the market model and explain the high frequency trading strategy. In Section 3, we formulate the control problem, derive the corresponding dynamic programming equation (DPE) for the value function,  and state some bounds and symmetry properties. We also simplify the DPE in two cases of practical interest, namely the case where the price is a martingale, and the case where the investor has predictive information on price trend available. In Section 4, we provide the numeri\-cal algorithm to solve the DPE, and we study the convergence of the numerical scheme towards the exact solution, by proving the monotonicity, stability and consistency for this scheme. We also provide numerical tests including computations of the optimal policies and performance analysis on a simulated data backtest. Finally, in Section 5, we show how to extend our model in the optimal liquidation case, i.e. when the investor's objective is to minimize the trading costs for  unwinding her portfolio.

\section{Market model}

\setcounter{equation}{0} \setcounter{Assumption}{0}
\setcounter{Theorem}{0} \setcounter{Proposition}{0}
\setcounter{Corollary}{0} \setcounter{Lemma}{0}
\setcounter{Definition}{0} \setcounter{Remark}{0}

Let us fix a probability space $(\Omega,\Fc,\Pb)$ equipped with a filtration $\F$ $=$ $(\Fc_t)_{t\geq 0}$ satisfying the usual conditions. It is assumed that all random variables and stochastic processes are defined on the stochastic basis $(\Omega,\Fc,\F,\Pb)$.

\vspace{1mm}

\noindent {\bf Prices in a one-tick microstructure.} We denote by $P$ the midprice, defined as a Markov process with generator $\Pc$ valued in $\P$. 
We shall assume that $P$ is a special semimartingale such that its predictable finite variation term $A$ satisfies the canonical structure: 
$dA_t$ $\ll$ $d<P>_t$, with a bounded density process:
\beq \label{theta}
\theta_t &=&  \frac{dA_t}{d<P>_t}, 
\enq
and the sharp  bracket process  $<P>$ is absolutely continuous with respect to the Lebesgue measure: 
\beq \label{vartheta}
d<P>_t &=& \varrho(P_t) dt, 
\enq
for some positive  continuous  function $\varrho$ on $P$.   We denote  by $\delta>0$  the tick size, and we shall assume that the spread is constantly equal to $\delta$, i.e. the best ask (resp. bid) price is $P^a := P + \frac{\delta}{2}$ (resp. $P^b := P - \frac{\delta}{2}$). 
This assumption corresponds to the case of the so-called \textit{one-tick microstructure} \cite{fielar08}, which can be encountered e.g. on short term interest rates futures contracts. 


\vspace{1mm}

\noindent {\bf Trading strategies.} For most of investors, the brokerage costs are paid when a transaction occurs, but new limit order submission, update or cancel are free of charge. Therefore, the investor can submit or update her quotes at any time, with no costs associated to this operation: it is then natural to model the limit order strategy (\textit{make} strategy) as a continuous time predictable control process. On the contrary, market orders lead to immediate execution, and are costly, so that continuous submission of market orders would lead to bankruptcy. Therefore, we choose to model the market order strategy (\textit{take} strategy) as impulse controls. More precisely, we model trading strategies by a pair $\alpha$ $=$ 
$(\alpha^{make},\alpha^{take})$ in the form:
\beqs
\alpha^{make} \; = \;  \big( L^a_t ,L^b_t \big)_{t \geq 0}, & &  \alpha^{take} \; = \;  \big( \tau_n , \xi_n \big)_{n \in \N}. 
\enqs
The predictable processes $L^a$ and $L^b$, valued in $\lbrace 0,1 \rbrace$ represent the possible \textit{make regimes}: when $L^a_t=1$ (resp. 
$L^b_t=1$) this means that the investor has active limit orders at the best ask price (resp. best bid price) at time $t$, else, if $L^a_t=0$ (resp. $L^b_t=0$) this means that the investor has no active order at the best ask price (resp. best bid price) at time $t$.  Practical implementation of such rule would be, for example, to send a limit order with a fixed quantity, 
when the corresponding control is $1$, and cancel it when it turns to $0$.  On the other hand, $(\tau_n)_{n\in\N}$ is an increasing sequence of stopping times, representing the times when the investor chooses to trade at market, and $\xi_n $, $n\geq 0$ are $\Fc_{\tau_n}$-measurable random variables valued in $\R$,  
representing the quantity purchased if $\xi_n\geq 0$ or sold if $\xi_n<0$. 

\vspace{1mm}

\begin{figure}[h!] 
\centering
\includegraphics[width=0.7\textwidth]{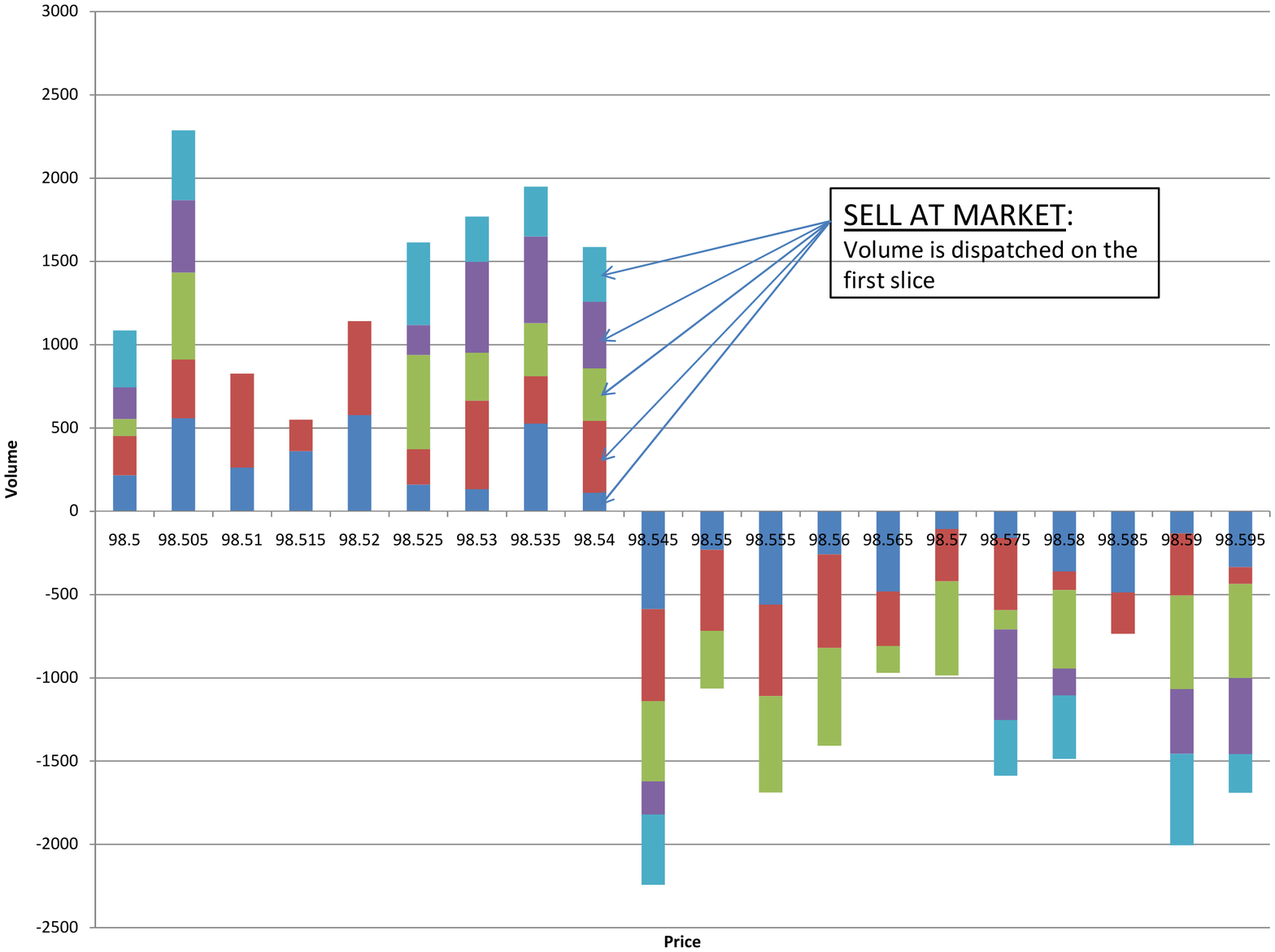} 
\caption{Schematic view of the pro-rata market microstructure.}
\label{prorata}
\end{figure}

\noindent {\bf Execution processes in a pro-rata microstructure.} The pro-rata microstructure (see \cite{jankab07} for extensive presentation and discussion) can be schematically described as follows\footnote{For a detailled description of actual trading rules, and a general overview of STIR futures trading, we refer to \cite{aik06} and references therein.}: when a market order comes in the pro-rata limit order book, its volume is dispatched among all active limit orders at best prices, proportionnally to each limit orders volumes, and therefore create several transactions (see Figure \ref{prorata}).

This pro-rata microstructure fundamentally differs from price-time microstructure \cite{guipha11} for two reasons: first, several limit orders at the best prices receive incoming  market order flow, 
regardless of the time priority,  and second, market makers tend to oversize their liquidity offering (that is, posting limit order with much higher volume than they actually want to trade) in order to increase their transaction volume. For example, on the three-months EURIBOR futures contracts, the liquidity available at the best prices is 200 times higher than the average transaction size. Therefore, to model a market making strategy, one must take into account the fact that limit orders are always oversized, so that  the executed volume is a random variable on which the market maker has no control\footnote{This differs from the price-time microstructure case, in which the market maker can control the upper bound of the execution volume by adjusting the limit order volume.}. More precisely, let  $N^a$  (resp. $N^b$) be a Poisson process of intensity $\lambda^a >0$ 
(resp. $\lambda^b$),  whose jump times represent the times when execution by a market order flow occurs at best ask  (resp. best bid), and we assume that  $N^a$  and  $N^b$ are independent.  
Let $(\zeta_n^a)_{n\in\N^*}$ and  $(\zeta_n^b)_{n\in\N^*}$ be two independent  sequences of i.i.d integrable random variables valued in 
$(0,\infty)$, of  distribution laws  $\mu^a$ and $\mu^b$, which represent the transacted volume of the $n^{th}$ execution at best ask and  best bid. 
We denote by $\nu^a(dt,dz)$ (resp. $\nu^b(dt,dz)$)  the Poisson random measure associated to the marked point process 
$(N^a,(\zeta_n^a)_{n\in\N^*})$ (resp.  $(N^b,(\zeta_n^b)_{n\in\N^*})$) of intensity measure $\lambda^a \mu^a(dz) dt$ 
(resp. $\lambda^b \mu^b(dz) dt$),  which is often identified with the compound Poisson processes
\beq
\label{execprocessesdynamics}
\vartheta_t^a \; = \;  \sum_{n=1}^{N^a_t} \zeta_n^a \; = \;  \int_0^t \int_0^\infty z \; \nu^a(dt,dz),  
& &  \vartheta_t^b \; = \;  \sum_{n=1}^{N^b_t} \zeta_n^b \; = \;  \int_0^t \int_0^\infty z \;  \nu^b(dt,dz). 
\enq
representing the cumulative volume of transaction at ask, and  bid.   
Notice that these processes model only the trades in which the investor has participated.

\vspace{1mm}

\noindent {\bf Cash holdings and inventory.}  The cash holdings process $X$ and the cumulated number of stocks  $Y$ (also 
called inventory) hold by the investor evolve according to the following dynamics:
\beq 
dX_t &=& L^a_t \big( P_{t^-} + \frac{\delta}{2}  \big) d\vartheta^a_t 
-  L^b_t \big( P_{t^-} -  \frac{\delta}{2}  \big) d\vartheta^b_t, \;\;\;\;  \tau_n \leq t < \tau_{n+1} \label{dynX}  \\
dY_t &=& L^b_t d\vartheta^b_t - L^a_t d\vartheta^a_t, \;\;\;\;\;\;\;  \tau_n \leq t < \tau_{n+1} \label{dynY} \\
X_{\tau_n} - X_{\tau_n-} &=&  - \xi_n P_{\tau_n} - \vert\xi_n \vert \big( \frac{\delta}{2}+\eps \big) -\eps_0,  \label{sautX} \\
Y_{\tau_n} -  Y_{\tau_n-} &=&   \xi_n.  \label{sautY}
\enq
The equations \reff{dynX}-\reff{dynY} model the evolution of the cash holdings and inventory 
under a limit order (make) strategy, while equations \reff{sautX}-\reff{sautY} describe the jump on the cash holdings and inventory when posting a market order (take) strategy, subject to a per share fee $\eps > 0$ and a fixed fee $\eps_0 >0$.  In the sequel, we  impose the natural admissibility condition 
that the size of the market order should not be larger than the current inventory, i.e. $|\xi_n|$ $\leq$ $|Y_{\tau_n-}|$, $n$ $\geq$ $0$, 
and we shall denote by $\Ac$ the set of  all admissible make and take strategies $\alpha$ $=$  $(\alpha^{make},\alpha^{take})$.

\vspace{1mm}

\begin{Remark}
{\rm  Let us define the process $V_t$ $=$  $X_t +Y_t P_t$, which represents at time $t$ the marked-to-market value of the portfolio (or book value of the portfolio).  From \reff{dynX}-\reff{dynY}-\reff{sautX}-\reff{sautY}, we see that its dynamics is governed by: 
\beq
\label{diffusion} dV_t &=& \frac{\delta}{2} ( L^b_t d\vartheta^b_t + L^a_t d\vartheta^a_t ) + Y_{t^-}  dP_t,  \\
\label{impulse} V_{\tau_n}-V_{\tau_n-} &=& - \vert \xi_n \vert (\frac{\delta}{2}+\varepsilon)-\eps_0.  
\enq
In equation \reff{impulse}, we notice that a trade at market will always diminish the marked to market value of our portfolio, due to the fact that we have to ``cross the spread", hence trade  at a least favorable price. On the other hand, in equation \reff{diffusion},  the term 
$\int \frac{\delta}{2}( L^b_t d\vartheta^b_t + L^a_t d\vartheta^a_t)$ is always positive, and represents the profit  obtained from  a limit order execution, while the  term $\int Y_{t^-} dP_t$ represents the portfolio value when holding  shares in the stock, hence inducing an inventory risk, which one wants to reduce its variance. }
 \end{Remark}
 

\section{Market making optimization procedure}

\setcounter{equation}{0} \setcounter{Assumption}{0}
\setcounter{Theorem}{0} \setcounter{Proposition}{0}
\setcounter{Corollary}{0} \setcounter{Lemma}{0}
\setcounter{Definition}{0} \setcounter{Remark}{0}

\subsection{Control problem formulation}
The market model in the previous section is fully determined by the state variables $(X,Y,P)$ controlled by the limit/market orders strategies 
$\alpha$ $=$ $(\alpha^{make},\alpha^{take})$ $\in$ $\Ac$. 
The market maker wants to optimize her profit over a finite time horizon $T$ (typically short term), while keeping control of her inventory risk, and to get rid of any risky asset by time $T$. We choose a  mean-variance optimization criterion, and the goal is to 
\beq \label{opti1}
\operatorname{ maximize } \;\;\;  \Eb \Big[ X_T  - \gamma \int_0^T Y_{t^-}^2 d<P>_t  \Big] 
\; \mbox{ over all strategies } \alpha \in \Ac, \; \mbox{ s.t} \;\;  Y_T = 0. 
\enq
The integral $\int_0^T Y_{t^-}^2 d<P>_t $ is a quadratic penalization  term for holding a non zero inventory in the stock, and  $\gamma$ $>$ $0$ is  a risk aversion parameter chosen by the investor.

Let us now rewrite problem \reff{opti1} in a more standard  formulation.  Notice indeed  that  
one can remove mathematically the constraint $Y_T=0$ on the inventory control, by introducing the liquidation function: 
\beqs
L(x,y,p) &=& x +  yp - \vert y \vert \big( \frac{\delta}{2} + \eps \big)-\eps_0,
\enqs
which represents the cash obtained after an immediate liquidation of the inventory via a market order. Thus, problem \reff{opti1} is formulated equivalently as
\beq \label{optimcriterion}
\operatorname{ maximize } \;\;\;  \Eb \Big[ L(X_T,Y_T,P_T) - \gamma  \int_0^T Y_t^2 \varrho(P_t)  dt \Big] 
\;\;\; \mbox{ over all strategies } \alpha \in \Ac,  
\enq
where we used also \reff{vartheta}. 
Let us then  define the value function for the problem \reff{optimcriterion}:
\beq \label{defvaluefunction}
v(t,x,y,p) &=&  \sup_{\alpha \in \Ac} \Eb_{t,x,y,p} \Big[ L(X_T,Y_T,P_T) - \gamma  \int_t^T Y_s^2 \varrho(P_s)  ds \Big],
\enq
for $t$ $\in$ $[0,T]$, $(x,y,p)$ $\in$ $\R^2\times\P$. Here, given $\alpha$ $\in$ $\Ac$, $\Eb_{t,x,y,p}$ denotes the expectation operator 
under which the process $(X,Y,P)$ solution to \reff{dynX}-\reff{dynY}-\reff{sautX}-\reff{sautY} with initial state $(X_{t^-},Y_{t^-},P_{t^-})$ $=$ $(x,y,p)$, 
is taken.  Problem \reff{defvaluefunction} is a mixed impulse/regular control problem in Markov model with jumps that we shall study by dynamic programming methods.

\vspace{1mm}

First, we state some bounds on the value function.

\begin{Proposition} \label{probound}
There exists some constant $K_{P}$ (depending only on the price process and $\gamma$) such that 
for all $(t,x,y,p)$ $\in$ $[0,T]\times\R^2\times\P$,
\beq \label{growth}
L(x,y,p)  \; \leq & v(t,x,y,p) & \leq \; x + yp  + \frac{\delta}{2}\big( \lambda^a \bar\mu^a + \lambda^b \bar\mu^b)(T-t) +  K_P, 
\enq 
where $\bar\mu^a$ $=$ $\int_0^\infty z \mu^a(dz)$,  $\bar\mu^b$ $=$ $\int_0^\infty z \mu^b(dz)$ are the mean of the distribution laws 
$\mu^a$ and $\mu^b$.
\end{Proposition}
{\bf Proof.}
The lower bound in \reff{growth} is derived easily  by considering the particular strategy, which consists of liquidating immediately  
all the current  inventory via a market order, and then doing nothing else until the final horizon.  Let us now focus on the upper bound.  Observe that  
in the definition of the value function in  \reff{defvaluefunction}, we can restrict obviously  to controls $\alpha$ $\in$ $\Ac$ s.t. 
\beq \label{integY}
\E \Big[ \int_0^T Y_t^2 d<P>_t \Big] & < & \infty.
\enq
For such strategies, we have: 
\beqs
& & \Eb_{t,x,y,p} \Big[ L(X_T,Y_T,P_T) - \gamma  \int_t^T Y_s^2 d<P>_s \Big] \\
& \leq &  \Eb_{t,x,y,p} \Big[ V_T - \gamma  \int_t^T Y_s^2 d<P>_s \Big]  \\
& \leq &  x + yp +   \Eb_{t,x,y,p} \Big[ \frac{\delta}{2}\big(\vartheta_{T-t}^a + \vartheta_{T-t}^b\big) + \int_t^T Y_{s^-} dP_s   - \gamma  \int_t^T Y_s^2 d<P>_s \Big] \\
&=& x + yp +   \Eb_{t,x,y,p} \Big[ \frac{\delta}{2}\big(\vartheta_{T-t}^a + \vartheta_{T-t}^b\big) + \int_t^T \big(Y_{s^-} \theta_s   - \gamma  Y_s^2 \big) d<P>_s \Big]. 
\enqs
Here, the second inequality follows from  the relation \reff{diffusion}, together with the fact that $L^a,L^b$ $\leq$  $1$,  $\vartheta^a$, $\vartheta^b$ are increasing processes, and also that jumps of $V$ are negative by \reff{impulse}.  The last equality holds true by \reff{theta} and the fact that 
$\int Y_- dM$ is a square-integrable martingale from \reff{integY}, where $M$ is the martingale part of the semimartingale $P$.  Since $\theta$ is bounded and $\gamma$ $>$ $0$, this shows that for all strategies $\alpha$ satisfying \reff{integY}, we have:
\beqs
& & \Eb_{t,x,y,p} \Big[ L(X_T,Y_T,P_T) - \gamma  \int_t^T Y_s^2 d<P>_s \Big] \\
& \leq & x + yp  +   \frac{\delta}{2} \E\big[ \vartheta_{T-t}^a + \vartheta_{T-t}^b] +   K \E[<P>_T],
\enqs
for some positive constant $K$, which proves the required result by recalling the characteristics of the compound Poisson processes $\vartheta^a$ and $\vartheta^b$, and since $<P>_T$ is assumed to be square-integrable.  
\ep

\vspace{2mm}

\begin{Remark}
{\rm The terms of the upper bound in \reff{growth} has a financial interpretation. The term $x+yp$ represents the marked-to-market value of the portfolio evaluated at mid-price, whereas the term $K_P$ stands for a bound on profit for any directional frictionless strategy on the fictive asset that is priced $P$. The term $\frac{\delta}{2}\big( \lambda^a \bar\mu^a + \lambda^b \bar\mu^b)(T-t)$, always positive, represents  the upper bound on profit due to market making, i.e. the profit of the strategy participating in every trade, but with no costs associated to managing its inventory.
}
\end{Remark}

\subsection{Dynamic programming equation}

For any $(\ell^a,\ell^b)\in \lbrace 0,1 \rbrace^2$, we introduce  the non-local operator associated with the limit order control:
\beq \label{defgenerator}
\Lc^{\ell^a,\ell^b}  &=&   \Pc  + \ell^a  \Gamma^a  + \ell^b\Gamma^b, 
\enq
where
\beqs
\label{defgammaa} 
\Gamma^a \phi (t,x,y,p) &=& \lambda^a \int_0^\infty \big[ \phi\big(t,x+ z(p +\dfrac{\delta}{2}),y-z, p\big)-\phi (t,x,y,p) \big]\mu^a(dz)  \\
\label{defgammab} 
\Gamma^b \phi (t,x,y,p) &=& \lambda^b \int_0^\infty \big[ \phi\big(t,x - z(p - \dfrac{\delta}{2}),y + z, p\big)-\phi (t,x,y,p) \big]\mu^b(dz),
\enqs
for $(t,x,y,p)$ $[0,T]\times\R\times\R\times\P$. 
Let us also consider the impulse operator associated with admissible  market order controls, and defined by:
\beqs
\Mc \phi(t,x,y,p) &=& \sup_{e \in [-|y|,|y|]} \phi\big(t,x-ep -\vert e \vert (\dfrac{\delta}{2}+\eps)-\eps_0,y+e,p\big). 
\enqs


The dynamic programming equation (DPE) associated to the control problem \reff{defvaluefunction} is a quasi-variational inequality (QVI) in the form: 
\beq \label{dynv}
\min\Big[ - \Dt{v} -  \sup_{(\ell^a,\ell^b)\in \lbrace 0,1 \rbrace^2}\Lc^{\ell^a,\ell^b} v \; + \;  \gamma g  \; , \; v - \Mc v \Big] &=& 0,   
\; \mbox{ on }  \;  [0,T)\times\R^2\times\P,
\enq
together with the terminal condition:
\beq \label{termv}
v(T,.) &=& L, \;\;\;  \; \mbox{ on }  \;  \R^2\times\P, 
\enq  
where we denoted by $g$ the function: $g(y,p)$ $=$ $y^2\varrho(p)$. This DPE  may be written explicitly as:
\beq 
\min\Big[ - \Dt{v} -  \Pc v -  \lambda^a \Big( \int_0^\infty \big[ v\big(t,x+ z(p +\dfrac{\delta}{2}),y-z, p\big)-v (t,x,y,p) \big]\mu^a(dz) \Big)_+ 
& & \label{dynv2} \\
\;\;\;\;\; - \;  \lambda^b \Big( \int_0^\infty \big[ v\big(t,x- z(p -\dfrac{\delta}{2}),y+z, p\big)-v (t,x,y,p) \big]\mu^b(dz) \Big)_+   + \gamma y^2\varrho(p) \; ; & &  \nonumber \\
\;\;\;\;\;\;\; v(t,x,y,p) -   \sup_{e \in [-|y|,|y|]} v\big(t,x-ep -\vert e \vert (\dfrac{\delta}{2}+\eps)-\eps_0,y+e,p\big) \Big] &=& 0,  \nonumber 
\enq
for $(t,x,y,p)$ $\in$ $[0,T)\times\R^2\times\P$, together with the terminal condition:
\beq \label{termv2}
v(T,x,y,p) &=& x+ yp - \vert y \vert \big( \frac{\delta}{2} + \eps \big)-\eps_0, \;\;\;  \forall (x,y,p) \in  \R^2\times\P. 
\enq  
By standard methods of dynamic programming, one can show that the value function in \reff{defvaluefunction} is the unique viscosity solution 
under  growth conditions determined by \reff{growth} to the   DPE \reff{dynv2}-\reff{termv2} of dimension $3$ (in addition to the time variable).

\subsection{Dimension reduction in the L\'evy case}

We now consider a special case on the mid-price process 
where the market making control problem can be reduced to the resolution of a one-dimensional 
variational inequality invol\-ving only the inventory state variable.  We shall suppose actually that $P$ is a L\'evy process so that
\beq \label{condcons}
\Pc I_{\P} \; = \;  c_{_P}, & \mbox{ and } & \varrho  \; \mbox{ is a constant}, 
\enq
where $I_\P$ is the identity function on $\P$, i.e. $I_\P(p)$ $=$ $p$, and $\varrho$ $>$ $0$, $c_{_P}$   are  real constants  depending on the 
characteristics triplet of $P$.  Two practical  examples are: 
 
\vspace{1mm}

\noindent $\bullet$ {\bf Martingale case}: The mid-price process $P$ is a martingale, so that $\Pc I_\P$ $=$ $0$.  This martingale assumption in a high-frequency context  reflects the idea that the market maker has no information on the future direction of the stock price. 

\vspace{1mm}

\noindent $\bullet$ {\bf  Trend information}:  To remove the martingale assumption, one can introduce some knowledge about the price trend.  
A typical simple example is when $P$ follows an arithmetic Brownian motion (Bachelier model). A more relevant example is described by 
a  pure jump process $P$ valued in the discrete grid $\delta \Z$ with tick $\delta$ $>$ $0$, and such that
\beqs
\Pb \big( P_{t+h}-P_t = \delta\,\vert\Fc_t \big) &=& \pi^+ h+o(h) \\
\Pb\big( P_{t+h}-P_t  = -\delta\,\vert\Fc_t\big)&=& \pi^- h+o(h)  \\
\Pb\big( \vert P_{t+h}-P_t \vert>\delta \,\vert\Fc_t\big)&=& o(h),
\enqs
where $\pi^+$, $\pi^-$ $>$ $0$, and $o(h)$ is the usual notation meaning that $\lim_{h\rightarrow 0} o(h)/h$ $=$ $0$.  Relation \reff{condcons} 
then holds with  $c_{_P}$ $=$ $\varpi \delta$, where $\varpi$ $=$ $\pi^+-\pi^-$ represents a constant information about price direction, and 
$\varrho$ $=$ $(\pi^+ + \pi^-)\delta^2$.  In a high-frequency context, this model is of practical interest as it provides a way to include a (predictive) information about price direction. For example, work have been done in \cite{conlar10} to infer the future prices movements (at the scale of a few seconds) from the current state of the limit order book in a Poisson framework. In this work, as well as in our real data tests, the main quantities of interest are the volume offered at the best prices in the limit order book, also known as the imbalance.

\vspace{2mm}

In this L\'evy context,  the value function $v$ is decomposed into the form: 
\beq \label{decvw}
v(t,x,y,p) &=& L(x,y,p)+ w(t,y), 
\enq 
where $w$ is solution to the integral  variational inequality:
\beq \label{QVIw}
\min\Big[  - \Dt{w} - y c_{_P} + \gamma \varrho y^2  - \Ic^a w  - \Ic^b w  \; , \; w - \tilde \Mc w \Big] &=& 0, \;  \mbox{ on } \;  [0,T)\times\R,
\enq
together with the terminal condition:
\beq \label{termw}
w(T,y) &=& 0, \;\;\;  \forall y \in  \R,
\enq
where $\Ic^a$ and $\Ic^b$ are the nonlocal integral operators:
\beqs
\Ic^a w(t,y) &=& \lambda^a \Big( \int_0^\infty \Big[ w(t,y-z)-w (t,y) + z \frac{\delta}{2} + (\frac{\delta}{2}+\eps)(|y| - \vert y-z\vert) \Big]\mu^a(dz) \Big)_+ \\
\Ic^b w(t,y) &=& \lambda^b \Big( \int_0^\infty \Big[ w(t,y+z)-w (t,y) + z \frac{\delta}{2}+ (\frac{\delta}{2}+\eps)(|y| - \vert y+z\vert) \Big]\mu^b(dz) \Big)_+, 
\enqs
and $\tilde\Mc$ is the nonlocal operator:
\beqs
\tilde\Mc w(t,y) &=&  \sup_{e \in [-|y|,|y|]} \Big[ w(t,y+e) - (\dfrac{\delta}{2}+\eps)(\vert y+e \vert + |e| -  \vert y \vert) -\eps_0 \Big]. 
\enqs
The interpretation of the decomposition \reff{decvw} is the following. The term $L(x,y,p)$ represents the book value that the investor would obtain by liquidating immediately with a market order, and $w$ is an additional correction term taking into account the illiquidity effects induced by the bid-ask spread and the fee, as well as the execution risk when submitting limit orders. Moreover, in the L\'evy case, this correction function $w$ depends only on time and  inventory.    
From \reff{growth}, we have the following bounds on the function $w$:
\beqs
0 \; \leq & w(t,y) & \leq (\frac{\delta}{2} +\eps)\vert y \vert + \frac{\delta}{2}\big( \lambda^a \bar\mu^a + \lambda^b \bar\mu^b)(T-t) +  K_P,
  \;\;\; \forall (t,y) \in [0,T]\times\R. 
\enqs
Actually,  we have a sharper upper bound in the L\'evy context.

\begin{Proposition} \label{supersolution}
Under \reff{condcons}, we have:
\beqs
0 \; \leq \;  w(t,y)  & \leq &  (T-t)\Big[ \dfrac{c_P^2}{4\gamma \rho}+ \lambda^a(\delta+\epsilon)\bar{\mu}^a + \lambda^b(\delta+\epsilon)\bar{\mu}^b\Big],
\enqs
for all  $(t,x,y,p) \in [0,T]\times\R^2\times\P$. 
\end{Proposition}
{\bf Proof.} 
For any $(x,y,p)$ $\in$ $\R^2\times\P$, we notice that
\beq
\nonumber &&L(x,y,p) - \sup_{e \in [-|y|,|y|]} L(x-ep -\vert e \vert (\dfrac{\delta}{2}+ \eps)-\eps_0,y+e,p)\\
\label{obstaclesuper} 
&=&  \eps_0 + (\dfrac{\delta}{2}+ \eps) \Big[ -\vert y \vert + \inf_{e \in [-|y|,|y|]} \vert e \vert + \vert y+e \vert\Big] \; = \;   \eps_0 > 0.
\enq
We also observe that  for all  $z\geq 0$:
\beq
L(x+z(p+\dfrac{\delta}{2}),y-z,p) -L(x,y,p)
 &=& z \dfrac{\delta}{2} + (\dfrac{\delta}{2}+ \eps)\big( \vert y \vert - \vert y-z \vert \big) \nonumber \\
 &\leq &   (\delta + \eps) z,  \label{diffsuper}
\enq
and similarly:
\beq
\label{diffsuper2} L(x-z(p-\dfrac{\delta}{2}),y+z,p) -L(x,y,p) & \leq & (\delta + \eps) z. 
\enq
Let us then consider the function $\phi (t,x,y,p)$ $=$  $L(x,y,p) +(T-t)u$, for some real constant $u$ to be determined later. Then, 
$\phi(T,.)$ $=$ $L$, and by  \reff{diffsuper}-\reff{diffsuper2}, we easily check that:
\beqs
&&-\dfrac{\partial \phi}{\partial t} - \sup_{(\ell^a,\ell^b)\in\{0,1\}^2} \mathcal{L}^{\ell^a,\ell^b} \phi \; + \; \gamma g \\
&\geq & u - \lambda^a (\delta + \eps) \bar{\mu}^a - \lambda^b (\delta + \eps) \bar{\mu}^b -y c_P +\gamma y^2 \rho.
\enqs
The r.h.s. of this last inequality  is a second order polynomial in $y$ and therefore it is always nonnegative  iff:
\beqs
c_P^2 - 4 \gamma \rho (u- \lambda^a (\delta + \eps) \bar{\mu}^a - \lambda^b (\delta + \eps) \bar{\mu}^b) &\leq& 0,
\enqs
which is satisfied once the constant $u$ is large enough, namely: 
\beqs
u & \geq & \hat u \; := \; \dfrac{c_P^2}{4\gamma \rho}+ \lambda^a(\delta+\epsilon)\bar{\mu}^a + \lambda^b(\delta+\epsilon)\bar{\mu}^b.
\enqs
For such choice of $u$ $=$ $\hat u$, and denoting by $\hat\phi$ the associated function:  $\hat\phi(t,x,y,p)$ $=$  $L(x,y,p) +(T-t)\hat u$ we have  
\beqs
-\dfrac{\partial \hat \phi}{\partial t} - \sup_{(\ell^a,\ell^b)\in \{ 0,1\}^2} \mathcal{L}^{\ell^a,\ell^b} \hat\phi + \gamma g &\geq &  0,
\enqs
which shows, together with \reff{obstaclesuper}, that $\hat \phi$ is a supersolution of \reff{dynv}-\reff{termv}. From comparison principle for this variational inequality, 
we deduce that
\beqs
v  & \leq & \hat \phi \,\, \text{ on } \,\, [0,T]\times\R^2\times\P,
\enqs
which shows the required upper bound for $w$ $=$ $v-L$. 
\ep

 \vspace{2mm}

Finally, from \reff{QVIw}-\reff{termw}, and in the case where $\lambda^a$ $=$ $\lambda^b$, $\mu^a$ $=$ $\mu^b$,  and by 
stressing the dependence of $w$ in $c_{_P}$, we see that $w$ satisfies the symmetry relation: 
\beq \label{symw}
w(t,y,c_{_P}) &=& w(t,-y,-c_{_P}), \;\;\; \forall (t,y) \in [0,T]\times\R. 
\enq

\section{Numerical  resolution}

\setcounter{equation}{0} \setcounter{Assumption}{0}
\setcounter{Theorem}{0} \setcounter{Proposition}{0}
\setcounter{Corollary}{0} \setcounter{Lemma}{0}
\setcounter{Definition}{0} \setcounter{Remark}{0}

In this section, we focus on the numerical resolution of the integral variational inequality \reff{QVIw}-\reff{termw}, which characterizes the 
reduced value function of the market-making problem in the L\'evy case.

\subsection{Numerical scheme}


We provide a computational scheme for the integral variational inequality \reff{QVIw}.  We first consider a time discretization of the interval 
$[0,T]$ with time step $h$ $=$ $T/N$ and a regular time grid $\mathbb{T}_N$ $=$ $\lbrace t_k=kh\, ,\, k=0,\ldots,N\rbrace$. 
Next, we discretize and localize the inventory state space on a finite regular grid: for any $M>0$ and $N_Y \in \N$, and denoting by $\Delta_Y = \dfrac{M}{N_Y}$, we set: 
\beqs
\mathbb{Y}_M &=& \big\{ y_i = i \Delta_Y,  \; i = -N_Y, \ldots,N_Y \big\}. 
\enqs
We denote by  $\operatorname{Proj}_M(y):= -M\vee(y\wedge M)$, and consider the discrete approximating distribution of $\mu^a$ and $\mu^b$, defined by:
\beqs
\hat{\mu}^a \; = \;  \sum_{i \in \Z^+ } \mu^a([i\Delta_Y;(i+1)\Delta_Y))\delta_{i\Delta_Y} \,\,\, ,\,\, \, & &
\hat{\mu}^b \; = \;   \sum_{i \in \Z^+ } \mu^b([i\Delta_Y;(i+1)\Delta_Y))\delta_{i\Delta_Y},
\enqs
with $\delta_x$ the Dirac measure at $x$.  We then introduce the operator associated to the explicit time-space discretization of the integral variational inequality \reff{QVIw}: 
for any   real-valued function $\varphi$ on $[0,T]\times \R$,  $t\in[0,T]$, and $y\in \R$, we define:
\beqs
\Sc^{h,\Delta_Y,M}(t,y,\varphi) &= & \max \Big[ \Tc^{h,\Delta_Y,M}(t,y,\varphi) \; ; \;   \tilde\Mc^{h,\Delta_Y,M}(t,y,\varphi) \Big],
\enqs
where 
\beqs
\Tc^{h,\Delta_Y,M}(t,y,\varphi) &=& \varphi(t,y)  - h\gamma\varrho y^2 + h y c_P  \\
& & \; + \;  \lambda^a h  \Big( \int_0^\infty \big[\varphi(t, \operatorname{Proj}_M(y-z)) - \varphi(t,y)\big] \hat{\mu}^a(dz)  \\
& & \;\;\;\;\;\;\;\;\;\;\;\;\;\; + \;  \int_0^\infty \big[ \frac{\delta}{2} z  + (\frac{\delta}{2}+\eps)(|y| - \vert y-z\vert) \big]\mu^a(dz) \Big)_+ \\
&& \; + \;  \lambda^b h  \Big( \int_0^\infty \big[\varphi(t, \operatorname{Proj}_M(y+z)) - \varphi(t,y) \big] \hat{\mu}^b(dz) \\
& & \;\;\;\;\;\;\;\;\;\;\;\;\;\;   + \;  \int_0^\infty \big[\frac{\delta}{2} z  + (\frac{\delta}{2}+\eps)(|y| - \vert y+z\vert )\big]\mu^b(dz) \Big)_+, 
\enqs
and 
\beq
& & \tilde\Mc^{h,\Delta_Y,M}(t,y,\varphi) \nonumber   \\
&=& \sup_{\tiny{e \in \mathbb{Y}_M\cap [-|y|,|y|]}} 
\big[\varphi(t,\operatorname{Proj}_M(y+e)) - (\dfrac{\delta}{2}+\eps)(\vert y +e \vert + \vert e \vert -\vert y\vert ) -\eps_0 \big].  \label{expressMc}
\enq
By recalling that $x_+$ $=$ $\max_{\ell \in \{0,1\}} \ell x$, we see that the operator $\Tc^{h,\Delta_Y,M}$ may be written also as:
\beq
\Tc^{h,\Delta_Y,M}(t,y,\varphi) &=& - h\gamma\varrho y^2 + h y c_P  +  
\max_{\ell^a,\ell^b \in \lbrace 0,1 \rbrace } \Big[ \varphi(t,y) (1-\lambda^a h \ell^a - \lambda^b h \ell^b) \label{expressLc} \\
& & \;\;\;\;\;\;\;\;\;\;  + \;  \lambda^a h \ell^a \Big( \int_0^\infty \varphi(t, \operatorname{Proj}_M(y-z))\hat{\mu}^a(dz)  \nonumber \\
& & \hspace{2.5cm} + \;  \int_0^\infty \big[\frac{\delta}{2} z  + (\frac{\delta}{2}+\eps)(|y| - \vert y-z\vert)\big]\mu^a(dz) \Big) \nonumber \\
&& \;\;\;\;\;\;\;\;\;\;   + \;  \lambda^b h \ell^b \Big( \int_0^\infty \varphi(t, \operatorname{Proj}_M(y+z))\hat{\mu}^b(dz) \nonumber \\
& & \hspace{2.5cm}   + \;  \int_0^\infty \big[\frac{\delta}{2} z  + (\frac{\delta}{2}+\eps)(|y| - \vert y+z\vert )\big]\mu^b(dz) \Big) \Big].   \nonumber
\enq
Note that  the integral terms involving $\varphi$ in $\Sc^{h,\Delta_Y,M}(t,y,\varphi)$  are in fact finite sums, and therefore are readily computable.
 We also assume, for simplicity sake, that the terms $\int_0^\infty \frac{\delta}{2} z + (\frac{\delta}{2}+\eps)(\vert y\vert - \vert y-z\vert)\mu^a(dz)$ and 
$\int_0^\infty \frac{\delta}{2} z  + (\frac{\delta}{2}+\eps)(\vert y\vert - \vert y+z\vert)\mu^b(dz)$ are exactly computable.

\vspace{1mm}

We then approximate the solution $w$ to \reff{QVIw}-\reff{termw} by the function $w^{h,\Delta_Y,M}$ on $\T_N\times\Y_M$ solution 
to the computational  scheme:
\beq
 w^{h,\Delta_Y,M}(t_N,.) &=&  0 \label{numschemeterm}  \\
w^{h,\Delta_Y,M}(t_k,y) &=&  \Sc^{h,\Delta_Y,M}(t_{k+1},y,w^{h,\Delta_Y,M}) \,\, , \,\, k = 0, \ldots, N-1 \, , \, y \in \mathbb{Y}_M. \label{numscheme} 
\enq

\subsection{Convergence of the numerical scheme}

In this section, we study the convergence of the numerical scheme \reff{numschemeterm}-\reff{numscheme} by showing the 
monotonicity, stability and consistency properties of this scheme. We denote by $C_b^1([0,T]\times\R)$ the set of  bounded 
continuously differentiable functions on $[0,T]\times\R$,  with bounded derivatives. 

\begin{Proposition} (Monotonicity)

\noindent For any $h>0$ s.t. $h < \dfrac{1}{\lambda^a+\lambda^b}$ the operator $\Sc^{h,\Delta_Y,M}$ is non-decreasing in $\varphi$, 
i.e. for any $(t,y)\in  [0,T] \times \R$ and any $\varphi,\psi \in C_b^1([0,T]\times\R)$ , s.t. $\varphi \leq \psi$ :
\beqs
\Sc^{h,\Delta_Y,M}(t,y,\varphi) &\leq&  \Sc^{h,\Delta_Y,M}(t,y,\psi)
\enqs
\end{Proposition}
{\bf Proof.} From the expression \reff{expressLc}, it is clear that  $\Tc^{h,\Delta_Y,M}(t,y,\varphi)$, and then also 
$\Sc^{h,\Delta_Y,M}(t,y,\varphi)$  is monotone in $\varphi$ once $1-\lambda^a h - \lambda^b h$ $>$ $0$. 
\ep

\vspace{2mm}

\begin{Proposition}(Stability)

\noindent For any $h,\Delta_Y,M >0$ there exists a unique solution $w^{h,\Delta_Y,M}$ to \reff{numschemeterm}-\reff{numscheme}, and the sequence $(w^{h,\Delta_Y,M})$ is uniformly bounded: for any $(t,y)\in \mathbb{T}_N \times \mathbb{Y}_M$,
\beqs
- \eps_0 \; \leq \;  w^{h,\Delta_Y,M} (t,y) &\leq& (T-t)\Big[ \dfrac{c_P^2}{4\gamma \rho}+ \lambda^a(\delta+\epsilon)\bar{\mu}^a + \lambda^b(\delta+\epsilon)\bar{\mu}^b\Big].
\enqs
\end{Proposition}
{\bf Proof.}
Existence and uniqueness of $w^{h,\Delta_Y,M}$  follows from the explicit backward scheme \reff{numschemeterm}-\reff{numscheme}. 
Let us now prove the uniform bounds.  We consider the function 
\beqs
\Psi^\star(t) &=&  (T-t)\left[ \dfrac{c_P^2}{4\gamma \rho}+ \lambda^a(\delta+\epsilon)\bar{\mu}^a + \lambda^b(\delta+\epsilon)\bar{\mu}^b\right]
\enqs
and notice that $\Psi^\star(t)$ $\geq$ $\Sc^{h,\Delta_Y,M}(t+h,y,\Psi^\star)$ by the same arguments  
as in Proposition \ref{supersolution}. Moreover, we have, by definition, $w^{h,\Delta_Y,M} (T,y)$ $=$ $\Psi^\star(T)$ $=$ $0$, and therefore, a direct recurrence from  \reff{numschemeterm}-\reff{numscheme} shows that $w^{h,\Delta_Y,M} (t,y)$ $\leq$ $\Psi^\star(t)$ for all 
$(t,y)\in \mathbb{T}_n \times \mathbb{Y}_M$, which is the required upper bound for $w^{h,\Delta_Y,M}$.  

On the other hand,   we notice  that $\Sc^{h,\Delta_Y,M}(t,0,\varphi)$ $\geq$ $\varphi(t,0)$ for any function $\varphi$ on $[0,T]\times\R$, and 
$t$ $\in$ $[0,T]$, by considering the``diffusive" part of the numerical scheme with the particular controls $\ell^a=\ell^b=0$. Therefore, since 
$w^{h,\Delta_Y,M} (T,0)=0$, we obtain by induction  on \reff{numschemeterm}-\reff{numscheme} that $w^{h,\Delta_Y,M} (t,0)\geq 0$ for any 
$t\in\mathbb{T}_N$. Finally, considering the obstacle part of the numerical scheme, with the particular control $e=-y$, shows that $w^{h,\Delta_Y,M} (t,y)\geq w^{h,\Delta_Y,M} (t,0)-\eps_0 \geq -\eps_0$ for any  $(t,y)\in \mathbb{T}_N \times \mathbb{Y}_M$, which proves the required lower bound for 
$w^{h,\Delta_Y,M}$.
\ep

\vspace{2mm}

\begin{Proposition}(Consistency)

\noindent For all $(t,y)\in [0,T)\times \R$ and $\varphi \in C_b^1([0,T]\times\R)$, we have
\beq
& & \lim_{\tiny{\begin{array}{c}(h,\Delta_Y,M) \rightarrow (0,0,\infty) \\ (t',y')\rightarrow (t,y)\end{array}}} 
\dfrac{1}{h} \Big[ \varphi(t',y')-\Tc^{h,\Delta_Y,M}(t'+h,y',\varphi) \Big] 
\label{diffusivecons} \\
&=&   - \Dt{\varphi}(t,y)  - y c_{_P} + \gamma \varrho y^2  - \Ic^a \varphi(t,y) - \Ic^b \varphi(t,y)  \nonumber 
\enq
and 
\beq \label{obstaclecons}
 \lim_{\tiny{\begin{array}{c}(h,\Delta_Y,M) \rightarrow (0,0,\infty) \\ (t',y')\rightarrow (t,y)\end{array} }} \tilde\Mc^{h,\Delta_Y,M}(t'+h,y',\varphi)  
&=&    \tilde\Mc \varphi(t,y)
\enq
\end{Proposition}
{\bf Proof.}  The consistency relation \reff{obstaclecons} follows from the continuity of the function $(t,y,e)$ $\rightarrow$ 
$\varphi(t,y+e) - (\dfrac{\delta}{2}+\eps)( \vert y+e \vert  + \vert e \vert - \vert y \vert)-\eps_0$.  On the other hand, we have for all $(t',y')$ 
$\in [0,T)\times\R$,
\beq
\dfrac{1}{h} \Big[ \varphi(t',y')-\Tc^{h,\Delta_Y,M}(t'+h,y',\varphi) \Big]  &=& \frac{1}{h} \big[ \varphi(t',y')-\varphi(t'+h,y')\big]  
- y' c_{_P} + \gamma \rho y'^2 \label{interdiff} \\
& &   - \;  \Ic_a^{h,\Delta_Y,M}(t'+h,y',\varphi) -  \Ic_b^{h,\Delta_Y,M}(t'+h,y',\varphi), \nonumber 
\enq 
where  
\beqs
 \Ic_a^{h,\Delta_Y,M}(t,y,\varphi) &=&  \lambda^a   \Big( \int_0^\infty \big[\varphi(t, \operatorname{Proj}_M(y-z)) - \varphi(t,y)\big] \hat{\mu}^a(dz)  \\
& & \;\;\;\;\;\;\;\;\;\;\;\;\;\; + \;  \int_0^\infty \big[ \frac{\delta}{2} z  + (\frac{\delta}{2}+\eps)(|y| - \vert y-z\vert) \big]\mu^a(dz) \Big)_+ \\
\Ic_a^{h,\Delta_Y,M}(t,y,\varphi) &=&   \lambda^b   \Big( \int_0^\infty \big[\varphi(t, \operatorname{Proj}_M(y+z)) - \varphi(t,y) \big] \hat{\mu}^b(dz) \\
& & \;\;\;\;\;\;\;\;\;\;\;\;\;\;   + \;  \int_0^\infty \big[\frac{\delta}{2} z  + (\frac{\delta}{2}+\eps)(|y| - \vert y+z\vert )\big]\mu^b(dz) \Big)_+. 
\enqs
The three first   terms of \reff{interdiff} converge trivially to $ - \Dt{\varphi}(t,y)  - y c_{_P} + \gamma \varrho y^2$ as $h$ goes to zero and 
$(t',y')$ goes to $(t,y)$. Hence, in order to get the consistency relation,  it remains to prove the convergence of 
$\Ic_a^{h,\Delta_Y,M}(t'+h,y',\varphi)$ to $\Ic^a \varphi(t,y)$ as $(h,\Delta_Y,M)$ goes to $(0,0,\infty)$, and $(t',y')$ goes to $(t,y)$ (an identical argument holds for $\Ic_b^{h,\Delta_Y,M}(t'+h,y',\varphi)$).  By writing that $|x_+-x'_+|$ $\leq$ $|x-x'|$, we have
\beqs
& & \Big| \Ic_a^{h,\Delta_Y,M}(t'+h,y',\varphi) - \Ic^a \varphi(t,y) \Big| \\
& \leq &  \lambda^a \big| \varphi(t'+h,y')-\varphi(t,y) \big|   \\
& & + \; \lambda^a \Big| \int_0^\infty \varphi(t'+h, \operatorname{Proj}_M(y'-z)) \hat{\mu}^a(dz) \;    - \;   \int_0^\infty \varphi(t,y-z) \mu^a(dz) \Big| \\
& \leq &  \lambda^a \big| \varphi(t'+h,y')-\varphi(t,y) \big|  \\
& & \; + \;  \lambda^a \Big| \int_0^{M+y'}  \varphi(t'+h, y'-z) \hat{\mu}^a(dz) \; - \;  \int_0^{M+y'} \varphi(t,y-z) \mu^a(dz) \Big|  \\
& & \; + \;  \lambda^a  \Big| \int_{M+y'}^\infty \varphi(t'+h, -M) \hat{\mu}^a(dz) \;    - \;   \int_{M+y'}^\infty \varphi(t,y-z) \mu^a(dz) \Big|  \\
& \leq & \lambda^a \big| \varphi(t'+h,y')-\varphi(t,y) \big|  \\
& &  \; + \;  \lambda^a  \int_0^{\infty} \big|  \varphi(t'+h, y'-\kappa(z))  \; - \;  \varphi(t,y-z) \big| \mu^a(dz)   \\
& & \; + \; 2 \lambda^a  \Vert \varphi \Vert_\infty \mu^a\big([M+y',\infty)\big),
\enqs 
where we denote by $\kappa(z)$ $=$ $\lfloor \dfrac{z}{\Delta_Y} \rfloor \Delta_Y$.  
Here $\lfloor z\rfloor$ denotes the largest integer smaller than $z$. 
Since the smooth function $\varphi$ has bounded derivatives, 
say bounded by $\Vert \varphi^{(1)} \Vert_\infty$,  it follows that
\beqs
\Big| \Ic_a^{h,\Delta_Y,M}(t'+h,y',\varphi) - \Ic^a \varphi(t,y) \Big| & \leq & 
\lambda^a \Vert \varphi^{(1)} \Vert_\infty  \big( h + 2|y'-y| + \Delta_Y \big) \\
& & \;\;\; + \; 2 \lambda^a  \Vert \varphi \Vert_\infty \mu^a\big([M+y',\infty)\big),
\enqs
which proves that 
\beqs
\lim_{\tiny{\begin{array}{c}(h,\Delta_Y,M) \rightarrow (0,0,\infty) \\ (t',y')\rightarrow (t,y)\end{array} }}  \Ic_a^{h,\Delta_Y,M}(t'+h,y',\varphi) 
&=& \Ic^a \varphi(t,y),
\enqs
hence completing the consistency relation \reff{diffusivecons}.  
\ep

\vspace{2mm}

\begin{Theorem} (Convergence) 

\noindent The solution $w^{h,\Delta_Y,M}$ to the numerical scheme (\reff{numschemeterm}-\reff{numscheme}) converges locally uniformly to $w$ on $[0,T) \times \R$, as $(h,\Delta_Y,M)$ goes  to $(0,0,\infty)$. 
\end{Theorem}
{\bf Proof.}
Given the above monotonicity, stability and consistency properties, the convergence of the sequence $(w^{h,\Delta_Y,M})$ towards $w$, which is the unique bounded viscosity solution to \reff{QVIw}-\reff{termw}, follows from \cite{barsou91}. We report the arguments for sake of completeness. 
From the stability property, the semi-relaxed limits:
\beqs
w_*(t,y) & = & \liminf_{\tiny{\begin{array}{c}(h,\Delta_Y,M) \rightarrow (0,0,\infty) \\ (t',y')\rightarrow (t,y)\end{array}}} w^{h,\Delta_Y,M}(t',y'), \\
w^*(t,y) & = & \limsup_{\tiny{\begin{array}{c}(h,\Delta_Y,M) \rightarrow (0,0,\infty) \\ (t',y')\rightarrow (t,y)\end{array}}} w^{h,\Delta_Y,M}(t',y'),
\enqs
are finite lower-semicontinuous and upper-semicontinuous functions on $[0,T]\times\R$, and inherit the boundedness of $(w^{h,\Delta_Y,M})$. 
We claim that $w_*$ are $w^*$ are respectively  viscosity super and subsolution of \reff{QVIw}-\reff{termw}.  Assuming for the moment that this claim is true, we obtain by the strong comparison principle for  \reff{QVIw}-\reff{termw} that $w^*$ $\leq$ $w_*$. Since the converse inequality is  obvious by  the very definition of $w_*$ and $w^*$, this shows  that $w_*$ $=$ $w^*$ $=$ $w$ is the unique bounded continuous viscosity solution to 
\reff{QVIw}-\reff{termw}, hence completing the proof of convergence. 

In the sequel, we prove the viscosity supersolution property of $w_*$ (a symmetric argument  for the viscosity subsolution property of $w^*$ holds true). Let $(\bar t,\bar y)$ $\in$ $[0,T)\times\R$ and $\varphi$ a test function in $C_b^1([0,T]\times\R)$ s.t. $(\bar t,\bar y)$ is a strict global minimimum point of $w_*-\varphi$. 
Then, one can find a sequence $(t_n',y_n')$ in $[0,T)\times\R$,  and a sequence $(h_n,\Delta_Y^n,M_n)$  such that:
\beqs
(t_n',y_n') \; \rightarrow \; (\bar t,\bar y), \;\;\; (h_n,\Delta_Y^n,M_n) \; \rightarrow \; (0,0,\infty), 
\;\;\; w^{h_n,\Delta_Y^n,M_n} \; \rightarrow \; w_*(\bar t,\bar y), \\
(t_n',y_n') \; \mbox{ is  a global minimum point of } w^{h_n,\Delta_Y^n,M_n} - \varphi. 
\enqs
Denoting by $\zeta_n$ $=$ $(w^{h_n,\Delta_Y^n,M_n} - \varphi)(t_n',y_n')$, we have $w^{h_n,\Delta_Y^n,M_n}$ $\geq$ $\varphi + \zeta_n$.  From the definition of the numerical scheme $\Sc^{h_n,\Delta_Y^n,M_n}$, and its monotonicity, we then get: 
\beqs
\zeta_n + \varphi(t_n',y_n') & = &  w^{h_n,\Delta_Y^n,M_n}(t_n',y_n') \\
&=& \Sc^{h_n,\Delta_Y^n,M_n}(t_n'+h_n,y_n',w^{h_n,\Delta_Y^n,M_n}) \\
& \geq &  \Sc^{h_n,\Delta_Y^n,M_n}(t_n'+h_n,y_n',\varphi + \zeta_n) \; = \; \Sc^{h_n,\Delta_Y^n,M_n}(t_n'+h_n,y_n',\varphi) + \zeta_n\\
&=& \max\Big[ \Tc^{h_n,\Delta_Y^n,M_n}(t_n'+h_n,y_n',\varphi)  \; , \; \tilde\Mc^{h_n,\Delta_Y^n,M_n}(t_n'+h_n,y_n',\varphi)\Big] + \zeta_n,
\enqs
which implies
\beqs
\min\Big[ \frac{\varphi(t_n',y_n') -  \Tc^{h_n,\Delta_Y^n,M_n}(t_n'+h_n,y_n',\varphi)}{h_n} \; , 
\; \varphi(t_n',y_n') -  \tilde\Mc^{h_n,\Delta_Y^n,M_n}(t_n'+h_n,y_n',\varphi)\Big]   \; \geq \;  0. 
\enqs
By the consistency properties \reff{diffusivecons}-\reff{obstaclecons}, and by sending $n$ to infinity in the above inequa\-lity, we obtain the required viscosity supersolution property:
\beqs
\min\Big[ - \Dt{\varphi}(\bar t,\bar y)  - \bar y c_{_P} + \gamma \varrho \bar y^2 - \Ic^a \varphi(\bar t,\bar y) - \Ic^b \varphi(\bar t,\bar y) \; , \;  
\varphi(\bar t,\bar y) - \tilde\Mc \varphi(\bar t,\bar y) \Big] & \geq & 0. 
\enqs
\ep

\subsection{Numerical tests}

In this section, we provide numerical results for the (reduced-form) value function and optimal policies in the martingale case and the trend information case, and a backtest on simulated data for the trend information case. Parameters for these numerical tests are shown in Figure 
\ref{tableparameters}.

\begin{figure}[h!] 
\centering
\subfloat[Market and risk parameters]{
\begin{tabular}{|l|l|}
\hline
Parameter & Value\\
\hline
$\delta$	&	$12.5$ EUR/contract\\
$\eps$	&	$1.05$ EUR/contract	\\
$\eps_0$ & 0 \\
$\lambda$	&	$0.05 s^{-1}$	\\
$\mu$ & $\operatorname{exp}(1/\bar{\mu})$\\
$\bar{\mu}$	&	$20$	contracts\\
$\gamma$	&	$2.5 .10^{-5}$	\\
$T$	&	100	s\\
\hline
\end{tabular}
}
\subfloat[Discretization parameters]{
\begin{tabular}{|l|l|}
\hline
Parameter & Value\\
\hline
$N_Y$	&	100	\\
$N_T$	&	500	\\
$N_\varpi$	&	20	\\
\hline
\end{tabular}
}
\caption{Parameters for numerical results.}
\label{tableparameters}
\end{figure}

This set of parameters are chosen to be consistent with calibration data on the front maturity for 3-months EURIBOR future, see for example 
\cite{fielar08}. Within this section, and for this set of parameters, we will denote by $w^h$ the value function and by $\alpha^\star$ the make/take strategy associated with the backward numerical scheme \reff{numschemeterm}-\reff{numscheme}. Given a generic controlled process $Z$ and a control $\alpha \in \Ac$, we will denote $Z^{\alpha}$ the process controlled by $\alpha$. Unless specified otherwise, such processes will be supposed to start at zero: typically, we assume that the investor starts from zero cash and zero inventory at date $t=0$ in the following numerical tests. Finally, we will write indifferently $w^h(t,y,c_P)$ or $w^h(t,y) := w^h(t,y,0)$ to either stress or omit the dependence in $c_P$.

\vspace{1mm} 

\noindent $\bullet$\textbf{ The martingale case}: in the martingale case, we performed the algorithm \reff{numschemeterm}-\reff{numscheme} with parameters shown above.  

\begin{figure}[h!] 
\centering
\includegraphics[width=0.6\textwidth]{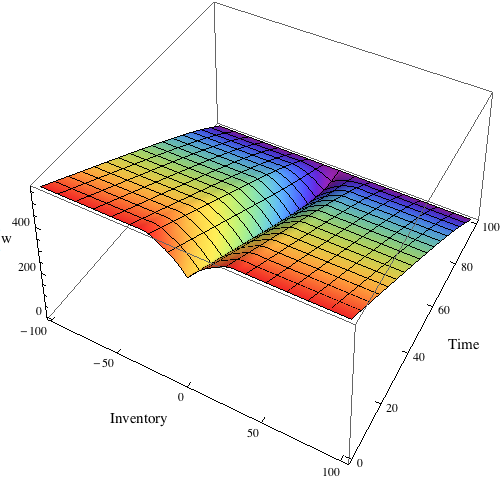} 
\caption{Reduced form value function $w^h$.}
\label{valfunmart}
\end{figure}

Figure \ref{valfunmart} displayed the reduced-form value function $w^h$ on $[0,T]\times [-N_Y;N_Y]$. This result illustrates the linear bound 
\reff{supersolution} as noticed in proposition \ref{supersolution}, and also the symmetry of $w^h$ as pointed out in \reff{symw}. We also observe 
the monotonicity over $\R^+$ and $\R^-$  of the value function $w^h(t,.)$.  

\begin{figure}[h!] 
\centering

\subfloat[Optimal take policy.]{
\includegraphics[width=0.47\textwidth]{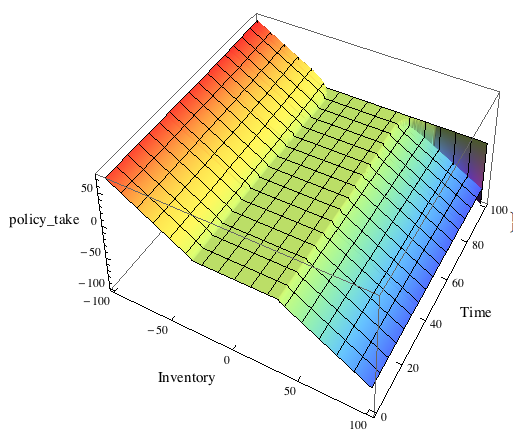} 
}
\subfloat[Optimal make policy]{
\includegraphics[width=0.53\textwidth]{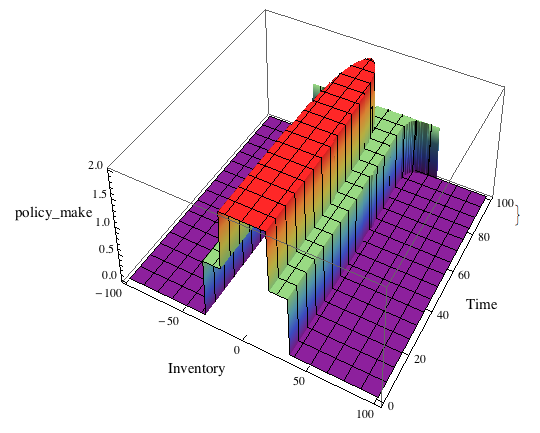} 
}
\caption{Numerical results for the martingale case: representation of optimal make and take policies $\alpha^\star$.}
\label{maketakemart}
\end{figure}

In Figure \ref{maketakemart},  we display the optimal make and take policies. The optimal take policy (on the left side) is represented as the volume to buy or sell with a market order, as a function of the time and inventory $(t,y)\in [0,T]\times [-N_Y;N_Y]$. 
We notice that a market order only occurs when the inventory becomes to large, and therefore, the take policy can be interpreted as a ``stop-loss" constraint, i.e. an emergency rebalancing of the portfolio when the inventory risk is too large.

The optimal make policy is represented as the regime of limit orders posting as a function of the time and inventory $(t,y)\in [0,T]\times [-N_Y;N_Y]$. For sake of simplicity, we represented the sum of $\ell^a$ and $\ell^b$ on the map. The meaning of this surface is as follows: $0$ means that there is no active limit orders on either sides, $2$ means that there is active limit orders on both bid and ask sides, and $1$ means that there is only one active limit order either on the bid or the ask side, depending on the sign of $y$ (if $y<0$ only the bid side is active, and if $y>0$ only the ask side is active). We notice that when close to maturity $T$, the optimal strategy tends to be more agressive, in the sense that it will seek to get rid of any positive or negative inventory, to match the terminal liquidation constraint. Moreover, we notice that close to date $0$, the dependence in $t$ seems to be negligible, which indicates that a``stationary regime" may be attained for large $T$.

 \vspace{3mm}

\noindent $\bullet$ \textbf{The trend information case:} in this case, we provide a backtest of the optimal strategy on simulated data in addition to the plots of the value function $w^h$ and optimal policy $\alpha^\star$. 

\begin{figure}[h!] 
\centering

\subfloat[Value function $w^h$ at date $t=0$.]{
\includegraphics[width=0.5\textwidth]{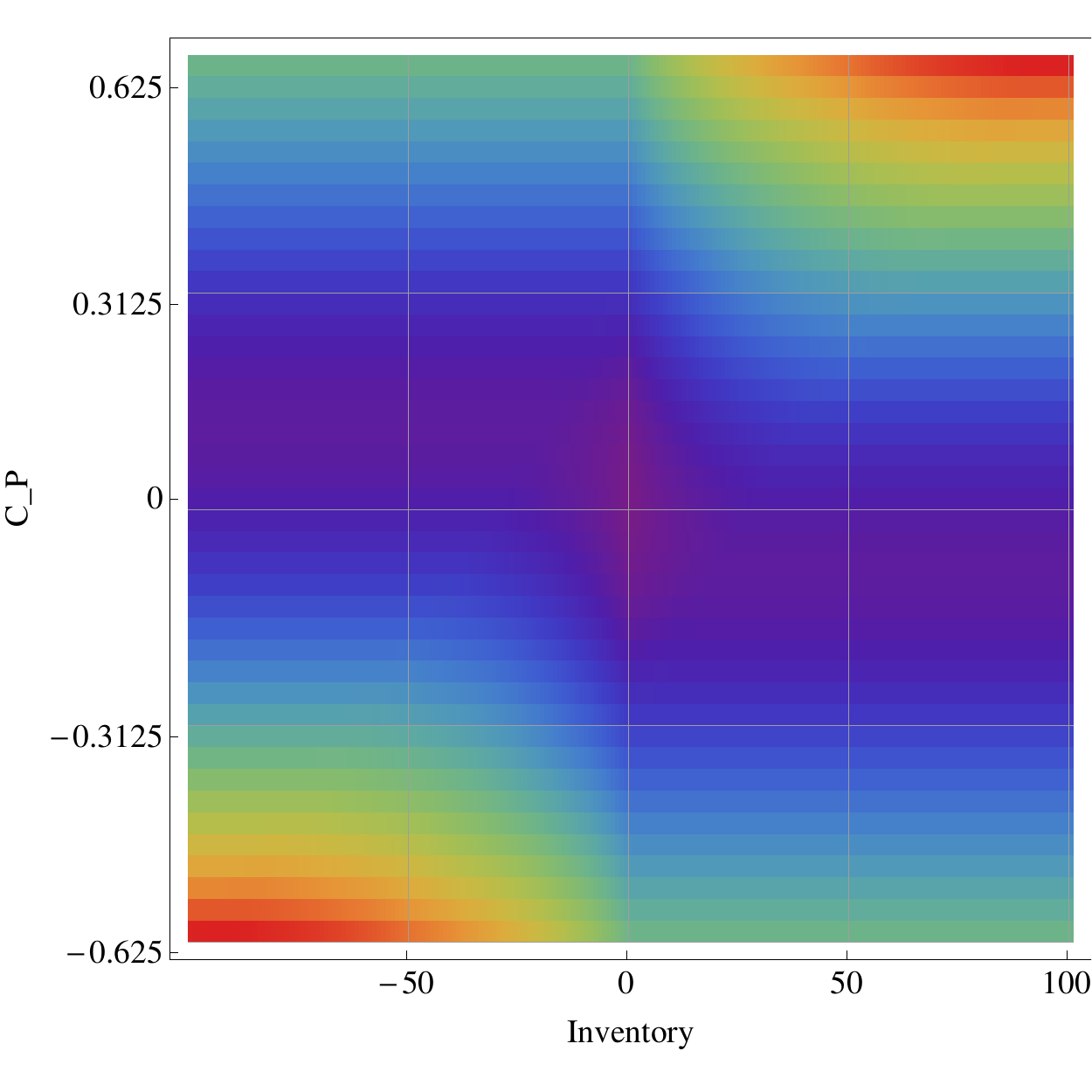} 
}
\subfloat[Optimal policy $\alpha^\star$ at date $t=0$.]{
\includegraphics[width=0.5\textwidth]{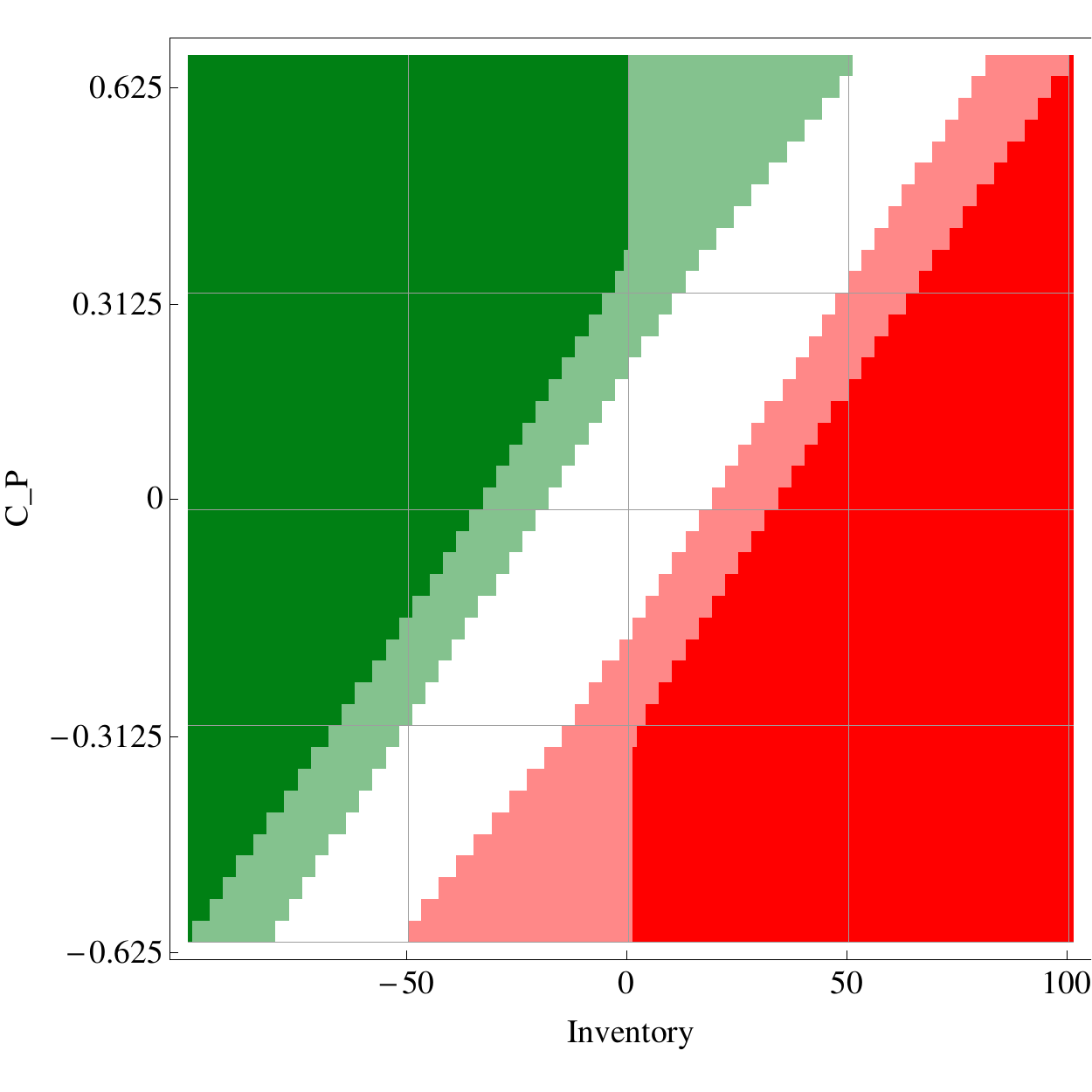} 
}
\caption{Value function and optimal policy for the trend information case.}
\label{trendinfo}
\end{figure}
Figure \ref{trendinfo} displays the value function and optimal policy at date $t=0$, in the plane $(y,c_P)$. The value function has central symmetry properties as expected in \reff{symw}, and should  be read as follows: dark green zones represent situation where a market order to buy must be sent, light green means that a limit order is active only at bid, white means that limit orders are active on both sides, light red means that a limit order is active only at ask, and dark red means that a market order to sell must be sent. The value function also increases with $\vert c_P \vert$. This effect can be interpreted as the gain in performance due to the superior information on price trend $c_P$. The interpretation of this extra performance due to $c_P$ is that the optimal policy avoids part of the adverse selection risk by using this predictive information about price movements. Let us provide a qualitative example:  assume that after the high frequency trader acquired a positive inventory, the adverse selection effect implies that price should go down; therefore, using the fact that in this case we should have $c_P<0$, the optimal strategy will be either to cancel the bid limit order (light red zone) and keep ask limit order active, or depending on the value of $\vert c_P \vert$, send a market order to get rid of our positive inventory (dark red zone).

We performed a benchmarked backtest on simulated data and a performance analysis in this case. The benchmark strategy is made of constant controls (a.k.a \textit{symmetric} or \textit{constant} strategy):
\beqs
\alpha^{benchmark}&:=&(\alpha^{make\, , \, benchmark},0)\\
\alpha^{make\, , \, benchmark} &:=& (1,1)
\enqs
In order to make our simulated data backtest closer to the reality, we chosed to slightly deviate from the original price model, and use a varying price trend. We simulate a price process model given by
\beqs
\hat{P}_t = \hat{P}_0 + \delta(N^+_t -N^-_t),
\enqs
where $N^+$ and $N^-$ are the Euler scheme simulation of Cox processes of respective intensities $\pi^+$ and $\pi^-$ defined as follows
\beqs
\pi^+ + \pi^- &\equiv& K = \varrho/\delta^2\\
d\pi^+_t - d\pi^-_t := d\varpi_t &=& -\theta \varpi_t dt + \sigma dB_t
\enqs
where $K>0$, $\theta>0$ and $\sigma>0$ are positive constants, and $B$ is an independent Brownian motion. Note that we choosed the sum $\pi^+ + \pi^-$ to be the constant $K$, for simplicity sake: it means that, disregarding the direction of price variation, the mean number of price change per second is assumed to be constant $\Pb \left( \vert P_{t+h}-P_t\vert = \delta \right)= Kh +o(h)$, which provides an easy way to calibrate the parameter $K$ while reducing the dimension of the simulation. The interpretation of this simulation model is as follows: we add an exogenous risk factor $B$, which drives the price trend information $\varpi$ as  an Ornstein-Uhlenbeck process. Notice that this supplementary risk factor $B$ is not taken into account in our optimization procedure and thus has a penalizing impact on the strategy's performance: therefore it does not spoil the backtest. This model choice for the process $(\varpi_t)$ is an convenient way to simulate the real-world situation, where the high-frequency trader continuously updates her predictive information about short-term price movements, based e.g. on the current state of the limit order book. 
Therefore, qualitatively speaking, our optimization procedure is consistent with this simulation model if we choose $\theta$ and $\sigma$ s.t.
the variation of the (reduced-form)  value function $w$ due to predictive information is very small compared to the variation of the value function due to other market events (e.g. an execution event). 
This assumption is consistent with HFT practice since the HF  trader is able to adapt very quickly to a modification of this predictive information. Backtest parameters involved in this simulation are shown in Figure \ref{backtest}.

\begin{figure}[h!] 
\centering
\begin{tabular}{|l|l|}
\hline
Parameter & Value\\
\hline 
$K$	&	1.0	\\
$\theta$	&	2	\\
$\sigma$	&	0.01	\\
$N_{MC}$	&	10000	\\
 \hline
\end{tabular}
\caption{Backtest parameters}
\label{backtest}
\end{figure} 

Let us  denote by $\hat{\vartheta}^a$ and $\hat{\vartheta}^b$ the Euler scheme simulation of the compound poisson processes $\vartheta^a$ and $\vartheta^b$, with dynamics \reff{execprocessesdynamics}. Therefore, for $\alpha\in\lbrace \alpha^\star, \alpha^{benchmark}\rbrace$, we were able to compute the Euler scheme simulation $\hat{X}^{\alpha}$ (resp. $\hat{Y}^{\alpha}$) of $X^{\alpha}$ (resp. $Y^{\alpha}$), starting at $0$ at $t=0$, by replacing $\vartheta^a$ (resp. $\vartheta^b$) by $\hat{\vartheta}^a$ (resp. $\hat{\vartheta}^b$) in equation \reff{dynX} (resp. \reff{dynY}).

We performed $N_{MC}$ simulation of the above processes. For each simulation $\omega\in[1...N_{MC}]$ and for $\alpha\in\lbrace \alpha^\star, \alpha^{benchmark}\rbrace$, we stored the following quantities: the terminal wealth after terminal liquidation $\hat{V}_T^{\alpha}(\omega):=L(\hat{X}^{\alpha}(\omega),\hat{Y}^{\alpha}(\omega),\hat{P}(\omega))$, called ``performance" in what follows ; the total executed volume $\hat{Q}^{total,\alpha}(\omega):=\sum_{[0,T]} \vert \hat{Y}^{\alpha}_t(\omega) - \hat{Y}^{\alpha}_{t-}(\omega)\vert$ ; and the volume executed at market $\hat{Q}^{market,\alpha}(\omega):=\sum_{[0,T]} \vert \xi_n(\omega)^{\alpha} \vert$. Finally, we denote by $m(.)$  the empirical mean, by $\Sigma(.)$ the empirical standard deviation, by $\operatorname{skew}(.)$  the empirical skewness, and  by  $\operatorname{kurt}(.)$ the empirical kurtosis, 
taken over $\omega\in[1...N_{MC}]$.  

\begin{figure}[h!] 
\centering
\begin{tabular}{|l|l|l|l|}
\hline
  Quantity & Definition & $\alpha^\star$ & $\alpha^{benchmark}$ \\
\hline \text{Info ratio over $T$}&$m(\hat{V}_T^.)/\sigma(\hat{V}_T^.)$ & 0.238 & 0.104 \\
\hline \text{Profit per trade} & $m(\hat{V}_T^.)/m(\hat{Q}^{total,.})$ & 1.37 & 3.86 \\
\text{Risk per trade} & $\sigma(\hat{V}_T^.)/m(\hat{Q}^{total,.})$ & 5.73 & 37.21 \\
 \hline \text{Mean performance}&$m(\hat{V}_T^.)$ & 376.08 & 773.15 \\
 \text{Standard deviation of perf}&$\sigma(\hat{V}_T^.)$ & 1574.97 & 7462.96 \\
 \text{Skew of perf}& $\operatorname{skew}(\hat{V}_T^.)$ & 0.027 & -0.0468 \\
 \text{Kurtosis of perf}& $\operatorname{kurt}(\hat{V}_T^.)$  & 3.21 & 7.48 \\
\hline 

\text{Mean total executed volume}&$m(\hat{Q}^{total,.})$ & 274.77 & 200.27 \\
 \text{Mean at market volume}&$m(\hat{Q}^{market,.})$ & 101.75 & 0. \\
 \text{Ratio market over total exec}&$m(\hat{Q}^{market,.})/m(\hat{Q}^{total,.})$ & 0.37 & 0.\\
 \hline
\end{tabular}
\caption{Synthetis table for backtest. Categories are, from top to bottom: relative performance metrics, period-adjusted performance metrics, absolute performance metrics and absolute activity metrics.}
\label{synthesis}
\end{figure}

Figure  \ref{synthesis} displayed a synthesis of descriptive statistics for this backtest. We first notice that the information ratio over $T$ of $\alpha^\star$ is more than twice that of $\alpha^{benchmark}$. Second, the per trade metrics can be compared to the half-spread $\dfrac{\delta}{2}=6.25 $ EUR/contract, and we see that although the mean profit per trade is smaller for the optimal strategy, the risk associated to each trade is dramatically reduced compared to the benchmark.\begin{figure}[h!] 
\centering
\includegraphics[width=0.7\textwidth]{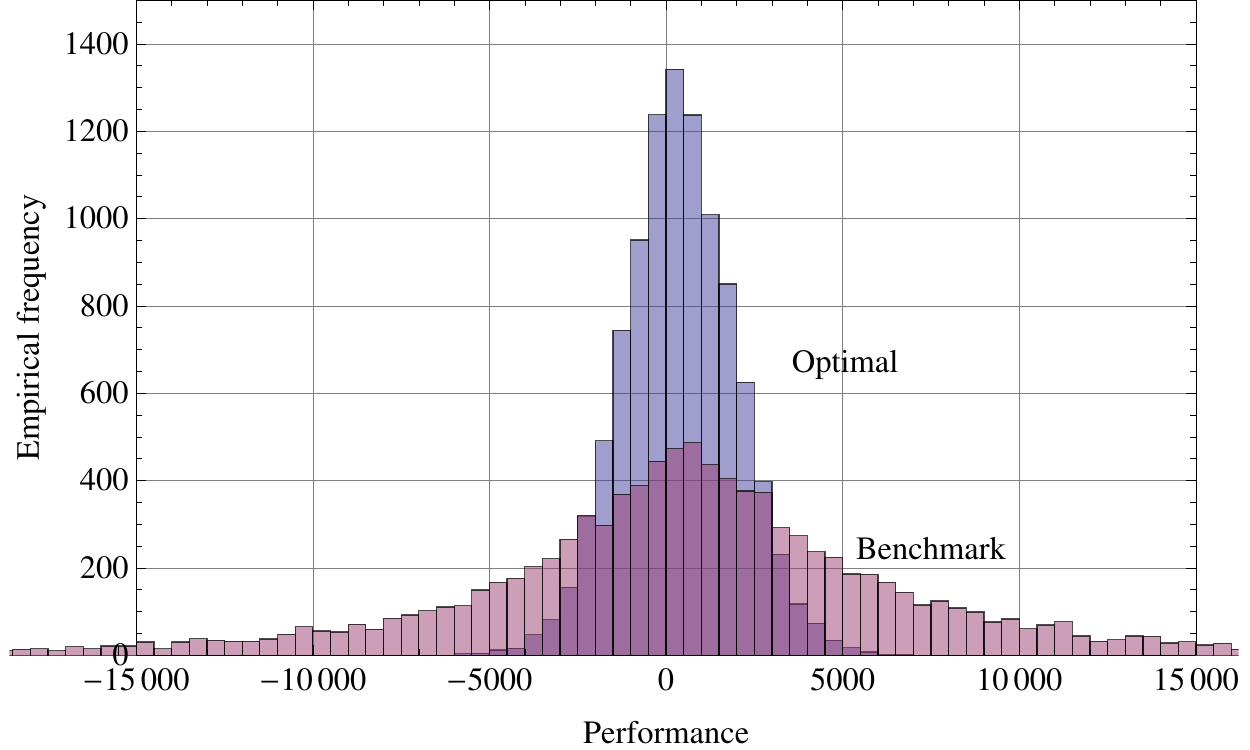} 
\caption{Empirical distribution of performance $\hat{V}_T^.$. The graph shows the number of occurences for each bin on $N_{MC}=10000$ simulations.}
\label{empdis}
\end{figure}
This is confirmed by the empirical distribution of performance, also shown in Figure \ref{empdis}, where the dark blue represents the performance distribution of the optimal strategy and the light purple represents the performance distribution of the benchmark strategy. We see that not only benchmark has higher standard deviation, but also higher excess kurtosis and heavy tails: this is due to the fact that inventory can be very large for the benchmark strategy, and therefore it bears a non-negligible market risk (or inventory risk). Finally, we see that about $37\%$ of the trades are done with market orders, which indicates that this feature of the strategy is relevant when exposed to adverse selection risk (the risk that the mid-price moves unfavourably after a limit order execution).

\begin{figure}[h!] 
\centering
\begin{tabular}{|l|l|l|}
\hline 
$\gamma$ & $\sigma(\hat{V}_T^.)$ & $m(\hat{V}_T^.)$ \\ 
\hline 
6.67.10{-04}	&	13.36	&	0.09	\\
4.44.10{-04}	&	351.16	&	20.98	\\
2.96.10{-04}	&	495.43	&	30.97	\\
1.98.10{-04}	&	649.11	&	39.28	\\
1.32.10{-04}	&	849.05	&	106.14	\\
8.78.10{-05}	&	1048.73	&	177.27	\\
5.85.10{-05}	&	1264.10	&	253.50	\\
3.90.10{-05}	&	1428.10	&	309.95	\\
2.60.10{-05}	&	1546.30	&	351.86	\\
1.73.10{-05}	&	1635.61	&	368.15	\\
1.16.10{-05}	&	1639.65	&	332.51	\\
\hline 
\end{tabular}
\caption{Varying risk aversion parameter $\gamma$: data.}
\label{efffrontdata}
\end{figure}

\begin{figure}[h!] 
\centering
\includegraphics[width=0.7\textwidth]{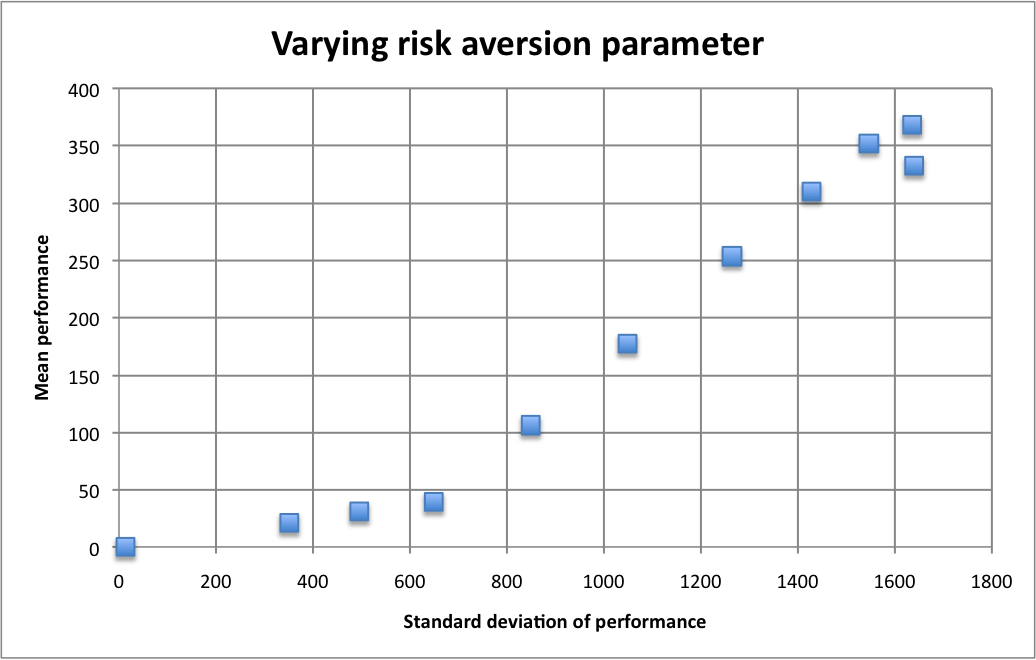} 
\caption{Varying risk aversion parameter $\gamma$: plot.}
\label{efffrontplot}
\end{figure}

Our last numerical test is devoted  to displaying the influence of the risk aversion para\-meter $\gamma$. All other parameters remaining the same, we tested several values of $\gamma$ (as indicated in Figure \ref{efffrontdata}), and characterized the performance of the corresponding strategy by the pair $(\sigma(\hat{V}_T^.),m(\hat{V}_T^.))$, which gives the \textit{efficient frontier} plot displayed in Figure \ref{efffrontplot}. As expected, a reduction of  $\gamma$  increases  the standard deviation of the strategy: this is due to the fact that a small $\gamma$ allows for large open position i.e. large inventory, and therefore the market risk is greater. For small $\gamma$, performance is also better since the investor can sustain large inventories, and therefore is less impatient to get rid of it: in particular, the proportion of volume executed at market is increasing in $\gamma$. We see that in real trading conditions, $\gamma$ must be chosen as small as possible, i.e. the value allowing the greatest performance, but maintaining the market risk sustainable.

\section{Best execution problem and  overtrading risk}

\setcounter{equation}{0} \setcounter{Assumption}{0}
\setcounter{Theorem}{0} \setcounter{Proposition}{0}
\setcounter{Corollary}{0} \setcounter{Lemma}{0}
\setcounter{Definition}{0} \setcounter{Remark}{0}

In this section, we apply our market model framework to a best execution problem.  The trading objective of the investor is to liquidate 
$Y_0>0$ assets over the finite time interval $[0,T]$. She is not allowed to purchase stock during the liquidation period, and may only buy back the asset in case of  short position.  In this context, the investor posts continuously a limit sell order (with a volume much larger that  the required quantity $Y_0$) at the best ask price, and also runs  market (sell) orders strategy until she reaches either a negative inventory or the terminal date. By doing so, she hopes to trade as much as possible at the ask price, and therefore avoiding to \textit{cross the spread}.  

Mathematically, this means  that the investor uses a subset  $\Ac_\ell$ of strategies $\alpha$ $=$ $(\alpha^{make}=(L^a,L^b),\alpha^{take})$ in $\Ac$ such that:
\beqs
(L_t^a,L_t^b)  &=& \left\{ \begin{array}{cc}
			   (1,0) & \mbox{ for } t  < \tau, \\
			   (0,0)  & \mbox{ for } t \geq \tau
			   \end{array}
			   \right.  \\
\alpha^{take} &=& (\tau_n,\zeta_n)_n \cup (\tau,-Y_\tau), \;\; \mbox{ with } \tau_n < \tau, \; \zeta_n < 0,			   
\enqs
where $\tau$ $=$ $\inf\{ t  \geq 0: Y_t \leq 0 \}$ $\wedge$ $T$. The value function associated to this liquidation problem is then defined by
\beq \label{defvaluefunction2}
v_\ell(t,x,y,p) &= & \sup_{\alpha \in  \Ac_\ell} \Eb_{t,x,y,p} \Big[ L(X_T,Y_T,P_T) - \gamma  \int_t^T Y_s^2 \varrho(P_s) ds \Big],
\enq
for $(t,x,y,p)$ $\in$ $[0,T]\times\R^2\times\P$.  With the notation in \reff{defgenerator}, the operator corresponding to the limit order in $\Ac_\ell$ is given by $\Lc^{1,0}$ $=$ $\Pc + \Gamma^a$, while the impulse operator associated to the market order in $\Ac_\ell$ is defined by:
\beqs
\Mc_\ell \; \varphi(t,x,y,p) & = & \sup_{e \in [-|y|,0]} \varphi\big(t,x-ep -\vert e \vert (\dfrac{\delta}{2}+\eps)-\eps_0,y+e,p\big)
\enqs
The dynamic programming equation associated to \reff{defvaluefunction2} takes the form: 
\beqs \label{dynv2}
\min\big[ - \Dt{v_\ell} -  \Pc v_\ell - \Gamma^a v_\ell  \; + \;  \gamma g  \; , \; v_\ell -  \Mc_\ell  v_\ell  \big] &=& 0,     \mbox{ on }   [0,T)\times\R\times (0,\infty)\times\P,
\enqs
together with the terminal and boundary conditions:
\beqs \label{termliq}
 v_\ell & = & L,  \;\;\;  \mbox{ on }   \big( \lbrace T \rbrace \times\R\times\R\times\P \big)\cup \big([0,T)\times\R\times\R_-\times\P\big).
\enqs
The above boundary  condition  for nonpositive inventory is related to  the \textit{overtrading risk}, which is  the risk that the investor sold too much assets via the (oversized) limit order at the best ask price. This risk occurs typically in  execution problems on pro-rata limit order book, see  \cite{fielar08}.

\vspace{1mm}

Again, in the L\'evy case \reff{condcons}, the value function $v_\ell$ is reduced into:
\beqs
v_\ell(t,x,y,p) &=& L(x,y,p) + w_\ell(t,y),
\enqs
where $w_\ell$ is solution to the integro-variational inequality:
\beqs
\min\Big\{ - \Dt{w_\ell} - y c_{_P}+ \gamma \varrho y^2   && \\
- \lambda^a  \int_0^\infty \Big[ w_\ell(t,y-z)-w_\ell(t,y) + z \frac{\delta}{2} + (\frac{\delta}{2}+\eps)(|y| - \vert y-z\vert) \Big]\mu^a(dz) \; ;  & & 
\label{QVIwl} \\
\;\;\;\;\;  w_\ell(t,y) -   \sup_{e \in [-y,0]} \Big[ w_\ell(t,y+e) - (\dfrac{\delta}{2}+\eps)(|y+e| + |e| - \vert y \vert) -\eps_0 \Big]  \Big\} &=& 0, \nonumber
\enqs
for $(t,y)$ $\in$ $[0,T)\times (0,\infty)$, together with the terminal and boundary conditions:
\beqs \label{termwl}
w_\ell(t,y) &=& 0, \;\;\;  \forall (t,y)  \in  \big(\{T\}\times\R\big) \cup \big( [0,T)\times\R_-\big).   
\enqs

\begin{figure}[h!] 
\centering
\subfloat[Value Function $w_\ell$]{
\includegraphics[width=0.5\textwidth]{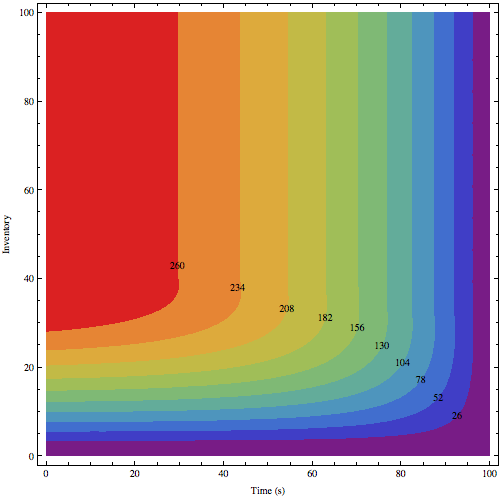} 
}
\subfloat[Optimal policy (take)]{
\includegraphics[width=0.5\textwidth]{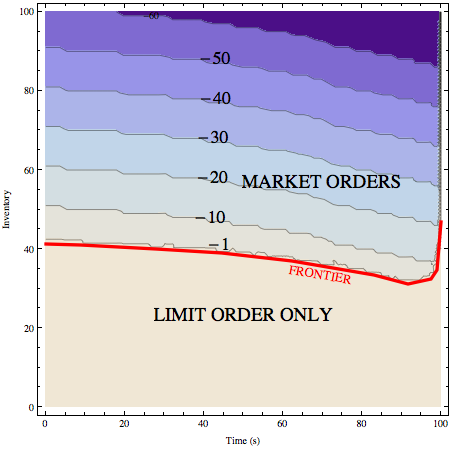} 
}
\caption{Numerical results for the simple liquidation problem (for $c_P = 0$). On the left side, level lines are indicated for the value function $w_\ell$. On the right side, numbers indicated on the figure represent the quantity to sell in the optimal market order control.}
\label{liq}
\end{figure}

The associated numerical scheme reads now as follows: 
\beqs
w_\ell^h(t_N,y) &=&   0, \;\;\;  y \in \R,   \\
w_\ell^h(t_k,y) &=&  0, \;\;\; k = 0,\ldots,N-1, \; y \leq 0, \\
w_\ell^h(t_k,y) &=&   \max \Big[ \Tc_\ell^{h,\Delta_Y,M}(t,y,\varphi) \; ; \;   \Mc_\ell^{h,\Delta_Y,M}(t,y,\varphi) \Big], 
\,\, k = 0, \ldots, N-1 \, , \, y \in \mathbb{Y}_M^+,  
\enqs
where  $\Y_M^+$ $=$ $\Y_M$ $\cap$ $\R_+$, 
\beqs
\Tc_\ell^{h,\Delta_Y,M}(t,y,\varphi) &=& \varphi(t,y)  - h\gamma\varrho y^2 + h y c_P  \\
& & \; + \;  \lambda^a h  \Big( \int_0^\infty \big[\varphi(t, \operatorname{Proj}_M(y-z)) - \varphi(t,y)\big] \hat{\mu}^a(dz)  \\
& & \;\;\;\;\;\;\;\;\;\;\;\;\;\; + \;  \int_0^\infty \big[ \frac{\delta}{2} z  + (\frac{\delta}{2}+\eps)(|y| - \vert y-z\vert) \big]\mu^a(dz) \Big) 
\enqs
and 
\beqs
& & \Mc_\ell^{h,\Delta_Y,M}(t,y,\varphi) \nonumber   \\
&=& \sup_{\tiny{e \in \mathbb{Y}_M \cap [-y,0]}} 
\big[\varphi(t,\operatorname{Proj}_M(y+e)) - (\dfrac{\delta}{2}+\eps)(\vert y +e \vert + \vert e \vert -\vert y\vert ) -\eps_0 \big].  
\enqs

In this case, the optimal policy shown in Figure \ref{liq} is simple to describe. 
 The state space is delimited in two zones: when the inventory is small, the HFT must wait for her limit sell order to be executed; and when the inventory is large, the HFT must send a market sell order to avoid the market risk related to holding a large position. 
 
 The frontier between the two zones (indicated in bold red in Figure \ref{liq}) can be interpreted as an \textit{optimal trading curve}, a concept that is extensively documented (see e.g. \cite{gueferleh11}) in the optimal execution literature.  The optimal trading curve is the inventory that the investor should hold, seen as a function of time, in order to minimize overall trading costs. Therefore, in the typical setting, the execution strategy consists in trading via market orders to get as close as possible to the optimal trading curve. Similarly, in our case, we can see on Figure \ref{liq} that the optimal strategy will behave similarly for large inventories (i.e. when above the trading curve): indeed, we observe that the quantities to sell are such that the market orders strategy would keep the inventory close to the optimal trading curve, if no limit orders were allowed. Now, in our case, we observe two specific features of the optimal strategy:
1) the optimal trading curve does not reach $0$ at maturity, and therefore the HFT has to get rid of her inventory at market at final date to match the constraint $Y_T=0$. This is due to the fact that a supplemental gain is always achievable when the limit order is executed. Therefore, this features leads to an execution strategy where the final trade is bigger than intermediary trades; 2) below the optimal trading curve, i.e. in the region where the HFT trades via limit orders only, the sell limit order is always active, and can lead to an execution. Therefore, the inventory is always below the optimal trading curve,
and the distance between the current inventory and the optimal trading curve equals the volume executed via limit orders. This differs from classic pattern-based best execution strategies, for example the U-shaped execution strategy that consists in trading a large quantity at the beginning and at the end of the liquidation, and trade regularly small quantities in between. Indeed, the optimal strategy does not provide a fixed pattern for every execution, but provide the optimal action to take given the observation of the inventory that is still to be sold and the market characteristics as e.g. the mean traded volume at ask per second $\lambda^a \bar{\mu}^a$, or trades volume distributions at ask $\mu^a$.

 

Finally, let us notice that this strategy can be interpreted as a convenient way to avoid the cost of crossing the spread during the liquidation of a portfolio, but we did not take into account the impact of the market order on the transaction price. In the case of a pro-rata microstructure, available volumes offered at best prices are usually about $200$ times larger  than the mean volume of market orders (see $\cite{fielar08}$), and therefore it is consistent to consider that there is no impact on the price for our market orders. Yet, the model can easily be modified by adding an impact component in the obstacle operator $\Mc_\ell$ to take care of this effect. We also did not model the possibility that the intensities $\lambda^a$ and 
$\lambda^b$ of execution processes may vary, and postpone this investigation  for future  research.


\vspace{9mm}

\begin{small}

\end{small}


\begin{thebibliography}{}

\bibitem{aik06} Aikin, S. (2006): ``Trading STIR Futures : An Introduction to Short-Term Interest Rate Futures", Harriman House Publishing, ISBN: 978-1897597811

\bibitem{avesto08} Avellaneda M. and S. Stoikov (2008): ``High frequency trading in a limit order book", 
{\it Quantitative Finance}, 8(3), 217-224.    

\bibitem{barsou91} Barles G. and P. Souganidis (1991): ``Convergence of approximation schemes for fully nonlinear second order equations", 
{\it Asymptotic Analysis}, 4, 271-283. 

\bibitem{carjairic11} Cartea A, Jaimungal S. and J. Ricci (2011): ``Buy low sell high: a high frequency trading perspective", preprint

\bibitem{conlar10} Cont R. and A. de Larrard (2010): ``Price Dynamics in a Markovian Limit Order Market", preprint

\bibitem{fielar08} Field J. and J. Large (2008): ``Pro-Rata Matching and One-Tick Futures Markets", preprint

\bibitem{gueferleh11} Gu\'eant O.,  Fernandez Tapia J. and  C.-A. Lehalle (2011): ``Dealing with inventory risk", preprint. 

\bibitem{LOBsurvey} Gould M.D., Porter M.A, Williams S., McDonald M., Fenn D.J. and S.D. Howison (2010): ``The limit order book: a survey", preprint.

\bibitem{guipha11} Guilbaud F. and H. Pham (2011): ``Optimal high frequency trading with limit and market orders", preprint.

\bibitem{jankab07} Karel Jane·cek, Martin Kabrhel (2007): ``Matching Algorithms of International Exchanges", working paper.


\bibitem{kuhstr10} K\"uhn C. and M. Stroh (2010): ``Optimal portfolios of a small investor in a limit order market: a shadow price approach", 
{\it Mathematics and Financial Economics}, 3(2), 45-72. 


\bibitem{ver11}  Veraart L.A.M. (2011): "Optimal Investment in the Foreign Exchange Market with Proportional Transaction Costs", 
{\it Quantitative Finance},  11(4): 631-640. 2011.
 

\end{thebibliography}
\end{document}